\documentclass[twocolumn]{aastex701}

\usepackage{amssymb}		
\usepackage{graphicx}		
\usepackage{hyperref}		
\usepackage{natbib}
\usepackage{float}

\usepackage{rotating}

\usepackage{amsmath}

\usepackage{soul}
\usepackage{silence}
\usepackage{perpage}
\MakePerPage{footnote}
\usepackage{lmodern}

\usepackage[table]{xcolor}

\DeclareRobustCommand{\ion}[2]{%
\relax\ifmmode
\ifx\testbx\f@series
{\mathbf{#1\,\mathsc{#2}}}\else
{\mathrm{#1\,\mathsc{#2}}}\fi
\else\textup{#1\,{\mdseries\textsc{#2}}}%
\fi}

\begin{document}

\title{Spectroscopic heterogeneity and multiscale ribbon fine structure during a solar flare observed by DKIST}

\author[0000-0002-3229-1848]{Cole A. Tamburri}
\email{c.tamburri@qub.ac.uk}
\affiliation{Astrophysics Research Centre, Queen's University Belfast,  Belfast, Northern Ireland, BT7 1NN, UK}

\author[0000-0001-8975-7605]{Maria D. Kazachenko}
\email{maria.kazachenko@lasp.colorado.edu}
\affiliation{National Solar Observatory, University of Colorado Boulder, 3665 Discovery Drive, Boulder, CO 80303, USA}
\affiliation{Department of Astrophysical and Planetary Sciences, University of Colorado Boulder, 2000 Colorado Ave, Boulder, CO 80305, USA}
\affiliation{Laboratory for Atmospheric and Space Physics, University of Colorado Boulder, 3665 Discovery Drive, Boulder, CO 80303, USA}

\author[0000-0001-9726-0738]{Ryan J. French}
\email{ryan.french@lasp.colorado.edu}
\affiliation{Laboratory for Atmospheric and Space Physics, University of Colorado Boulder, 3665 Discovery Drive, Boulder, CO 80303, USA}

\author[0000-0001-7458-1176]{Adam F. Kowalski}
\email{adam.f.kowalski@colorado.edu}
\affiliation{National Solar Observatory, University of Colorado Boulder, 3665 Discovery Drive, Boulder, CO 80303, USA}
\affiliation{Department of Astrophysical and Planetary Sciences, University of Colorado Boulder, 2000 Colorado Ave, Boulder, CO 80305, USA}
\affiliation{Laboratory for Atmospheric and Space Physics, University of Colorado Boulder, 3665 Discovery Drive, Boulder, CO 80303, USA}

\author[0000-0001-7949-9163]{Sarah Riley}
\email{sarah.riley1@montana.edu}
\affiliation{Department of Physics, Montana State University, Bozeman, MT 59717, USA}

\author[0000-0001-5316-914X]{Graham S. Kerr}
\email{graham.kerr.2@glasgow.ac.uk}
\affiliation{SUPA School of Physics and Astronomy, University of Glasgow, Glasgow, G12 8QQ, UK}
\affiliation{Department of Physics, Catholic University of America, 620 Michigan Avenue Northeast, Washington, DC 20064, USA}

\author[0000-0003-4065-0078]{Rahul Yadav}
\email{rahul.yadav@lasp.colorado.edu}
\affiliation{National Solar Observatory, University of Colorado Boulder, 3665 Discovery Drive, Boulder, CO 80303, USA}
\affiliation{Laboratory for Atmospheric and Space Physics, University of Colorado Boulder, 3665 Discovery Drive, Boulder, CO 80303, USA}

\author[0000-0003-1597-0184]{Marcel Corchado-Albelo}
\email{marcel.corchado@colorado.edu}
\affiliation{National Solar Observatory, University of Colorado Boulder, 3665 Discovery Drive, Boulder, CO 80303, USA}
\affiliation{Department of Astrophysical and Planetary Sciences, University of Colorado Boulder, 2000 Colorado Ave, Boulder, CO 80305, USA}
\affiliation{Laboratory for Atmospheric and Space Physics, University of Colorado Boulder, 3665 Discovery Drive, Boulder, CO 80303, USA}

\author[0000-0002-2797-744X]{Jiong Qiu}
\email{qiu@montana.edu}
\affiliation{Department of Physics, Montana State University, Bozeman, MT 59717, USA}

\begin{abstract}

On 11 August 2024, the Daniel K. Inouye Solar Telescope collected high-resolution, high-cadence observations of a C4.7-class solar flare, including Ca \textsc{II} H, H$\epsilon$, and H$\beta$ spectra from the Visible Spectropolarimeter (ViSP) and H$\alpha$ images from the Visible Broadband Imager (VBI).  Using an unsupervised machine learning technique to cluster the complex ViSP spectral profiles, we analyze the spectrotemporal evolution of flare ribbon substructure.  Our analysis reveals significant spatial variation in the spectroscopic behavior of the flare ribbon, notably four categories of spectral type: (i) redshifts with velocities up to $v_z\approx$ 22 $\text{km}\;\text{s}^{-1}$, (ii) unshifted, symmetric line profiles, (iii) central reversals, and (iv) blueshifts with velocities reaching $v_z\approx-$11$\;\text{km}\;\text{s}^{-1}$. Concurrently, H$\alpha$ images reveal that at times during the flare impulsive phase, part of the ribbon is composed of $\sim$300--470 km wide bead-like structures that later appear to splinter into 40--60 km wide loop-like features. Crucially, the number of observed splinters is correlated with the quality of observing conditions, suggesting that bead-like ribbon structures may generally consist of smaller fragments that are only visible under the best observing conditions. ViSP spectroscopy demonstrates that the ribbon fragments are characterized by enhancements in the red wing of Ca \textsc{II} H, indicative of plasma downflows.  Our analysis demonstrates the remarkable spectral diversity in the flare chromosphere and relates two recently discovered spatial regimes of solar flare fine structure: loop-like features tens of kilometers wide and ribbon substructure a few hundred kilometers wide.

\end{abstract}

\section{Introduction}\label{sec:introc6}

Solar flares are characterized by the release of an enormous amount of magnetic energy -- up to $10^{32}$ ergs -- on timescales from tens of seconds to several hours. The standard 2D flare model \citep{carmichael1963,sturrock1966,hirayama1974,kopp_pneuman_1976} dictates that magnetic reconnection in the flaring corona initiates the conversion of free magnetic energy into kinetic, thermal, and nonthermal particle energy. Much of this is deposited into the lower layers of the solar atmosphere, producing localized, brilliant \textit{ribbons} of enhanced emission at ultraviolet (UV) and extreme-ultraviolet (EUV) wavelengths and in a large number of prominent lines in the solar spectrum.  While the phenomenological reconnection-flare model guides our general understanding of flares on megameter scales, advances in observational and modeling capabilities have recently revealed the significance of fine-scale flare structure in the solar chromosphere and corona at spatial scales from hundreds of kilometers down to tens of kilometers (e.g., \citealt{pietrow2024,faber2025,yadav2025,french2025_beads,tamburri2025,faber2026,collier_2026}).  

Many aspects of flare fine structure remain unexplained, including the prevalence of periodic \textit{knots}, \textit{beads}, or \textit{blobs}, and extended \textit{riblets} that appear within flare ribbons when observed at high spatial resolution (e.g., \citealt{li_zhang2015,yadav2025,faber2025,french2025_beads}).  \cite{faber2026} proposed that these may be the same structures viewed from different perspectives: compact, circular, $\sim$110--360 km wide beads when observed at heliocentric viewing angles of $\mu=\cos\theta\approx1$ and extended $\sim$280--1500 km long riblets when observed at smaller $\mu$.  

Fragmented flare ribbon structure might be generated in part by reconnection dynamics such as the tearing instability (e.g., \citealt{wyper_pontin_2021,dahlin_2025,french2025_beads,ashfield2025_natas,corchado_albelo2026,chouliaras2026}).  In this scenario, physical processes in the current sheet imprint upon the chromosphere by modulating the rate and spatial structure of energy transfer. Alternative origins for periodic ribbon structure have been proposed, including the anatomy of the pre-flare chromosphere or the presence of ``uncombed'' chromospheric loops interspersed between actively flaring loops \citep{chitta2026}. Recent modeling work also suggests that coronal processes other than reconnection, such as supra-arcade downflows (SADs), may affect ribbon substructure by regulating the pattern of particle precipitation along lower-lying coronal loops \citep[e.g.,][]{li2025}. 

In the corona, observations from the Daniel K. Inouye Solar Telescope (DKIST; \citealt{rimmele2020,rast2021}) have revealed post-reconnection flare loops as small as $\sim$21 km, the diffraction limit of the Visible Broadband Imager (VBI; \citealt{woger2021}), with average widths of 45--55 km \citep{tamburri2025}. In the past, observations from the High-Resolution Coronal Imager \citep[Hi-C;][]{kobayashi_2014,peter2013,klimchuk2020,williams_evidence_2020,williams_is_2020,williams_cross-sectional_2021}, the Swedish Solar Telescope \citep[SST;][]{scharmer2003,antolin2012}; the Goode Solar Telescope \citep[GST;][]{jing2016,schmidt2025}, and Solar Orbiter \citep{muller2020,hayes2026} have measured coronal loops (both post-flare reconnection loops and non-flare active region loops) 100 km wide or larger, the same scale as the ribbon beads discussed above.  Crucially, the literature almost always reports loop widths at or below the diffraction limit of the observing instrument, complicating a characterization of fundamental scales in the flare atmosphere.

This presents a novel conundrum:  Are chromospheric ribbon beads composed of finer structures at scales of tens of kilometers or smaller, as might be expected from the magnetic-field gradient between the flaring chromosphere and corona? Alternatively, do the different layers of the solar atmosphere have distinct characteristic spatial scales, such that fundamental chromospheric scales are slightly larger than their coronal counterparts?

To augment detailed analysis of spatial structure in the flare atmosphere, high-resolution spectra from the SST, the Interface Region Imaging Spectrometer (IRIS, \citealt{depontieu2014}), Solar Orbiter, DKIST, and others have revealed diverse small-scale spectral behavior, challenging a simple model of uniform heating in the flare atmosphere.  For example, \cite{polito_2023} and \cite{kerr_2024} investigated spectral signatures unique to the leading edge of the flare ribbon front. These include dimming of the He \textsc{i} 10830~\AA\ line prior to the brightening that is observed in much of the main ribbon, and properties of \ion{Mg}{ii} near-UV spectra that suggest mid-chromospheric heating. These features were explainable by flare models only if the lower atmosphere was initially heated by a very low flux of very energetic electrons, with subsequent precipitation of a large energy flux (some 2-3 orders of magnitude larger) more typical of flares. 

In another work, \cite{pietrow2024} noted the spectral variability in the chromosphere during a large flare observed by SST. Spectral complexity is also evident in a flare captured by DKIST during the decay phase \citep{tamburri2026}, although this observation was limited by suboptimal observing conditions and missed the flare impulsive phase. \cite{yadav2025} investigated the spectral properties of fine structures in a small flare observed by DKIST and inverted the thermodynamic properties of the flare atmosphere, but were limited by relatively few scans and low temporal cadence such that the evolution of fine structures was unresolved.  Notably, significant spatial variations in flare spectra are not always seen. For example, \cite{graham_cauzzi_2015} found remarkably uniform evolution of \ion{Fe}{xxi} 1354~\AA\ and \ion{Mg}{ii} 2792~\AA\ Doppler shifts in an X-class flare ribbon observed by IRIS.

If observed at sufficiently high resolution, chromospheric flare spectra can be used to infer the physical mechanisms driving individual heating events within a flare ribbon. In one instance, spatial differences in spectral behavior could be explained by spatiotemporal variations in the energetic properties of the non-thermal electron beam that heats the lower atmosphere; in another, discrepant spectral signatures could be due to variation of the energy transport mechanism itself \citep[e.g.][and references therein]{kerr2023r,kerr2026}. These other mechanisms include beams of non-thermal protons \citep[e.g.][]{1995SSRv...73..387S,2009ApJ...698L.152S,kerr_2023_oz,sadykov2023}, direct heating of the corona \citep[e.g.][]{1991A&A...241..618G,longcope2014,polito_2023,lorincik_2025,kerr2026}, and Alfv\'en waves \citep[e.g.][]{emslie_sturrock1982,fletcher_hudson2008, reep_2016,kerr_2016,reep2018b,russell2024,lorincik_2025}. For example, \cite{kerr2026} recently discovered that two ribbon footpoints during a small flare observed by Solar Orbiter were likely heated by two entirely different mechanisms: one by a beam of electrons and one by the conductive flux from a directly-heated corona.  

The parameter space that describes potential flare heating mechanisms is evidently vast and largely unconstrained, and the physics governing flare fine structure is not yet clarified.  The coupling between spectral and spatial differences in flare fine structure motivates simultaneous investigation of high-resolution flare spectra and images in order to determine the individual heating scenarios responsible for flare emission and resolve the fundamental structures that make up the flaring atmosphere. As the world's largest solar telescope, the 4-meter aperture DKIST is uniquely capable of resolving scales down to $\sim$20~km with the VBI while simultaneously capturing high-resolution spectra with the Visible Spectropolarimeter (ViSP, \citealt{dewijn2022}).  DKIST flare spectroscopy and imaging are therefore invaluable to detailed studies of flare heating, but very few flares have been observed with the DKIST so far, especially during the impulsive flare phase. In this work we demonstrate that comprehensive observations of a solar flare with DKIST can be used to investigate both the spectral and spatial properties of flare fine structure.

In Section \ref{sec:obs_c6} we present DKIST observations of the Geostationary Operational Environmental Satellite (GOES) X-ray class C4.7 flare on 11 August 2024.  In Section \ref{sec:ViSPvar} we use an unsupervised machine learning technique to characterize diverse, spatially non-uniform chromospheric ViSP spectra in the flare ribbons, highlighting the complexity of impulsive phase heating. In Section \ref{sec:finestructure} we analyze the flare ribbon details in VBI H$\alpha$ images and ViSP spectra, bridging the divide between two spatial regimes of flare fine structure: the $\sim$45--55 kilometer wide coronal loops reported by \cite{tamburri2025} and the $\sim$150-500 kilometer wide ribbon beads investigated by \cite{yadav2025}, \cite{french2025_beads}, and \cite{faber2025}, among others. Crucially, many of the small-scale dynamics observed are currently resolvable only by the suite of instruments at DKIST. In Section \ref{sec:disc_c6} we discuss our analysis in the context of recent work.  Finally, in Section \ref{sec:concc6} we summarize our main findings.

\section{Observations}\label{sec:obs_c6}

\subsection{Summary of event}\label{sec:summary}

\begin{figure*}
     \begin{center}
    \includegraphics[width=1\textwidth]{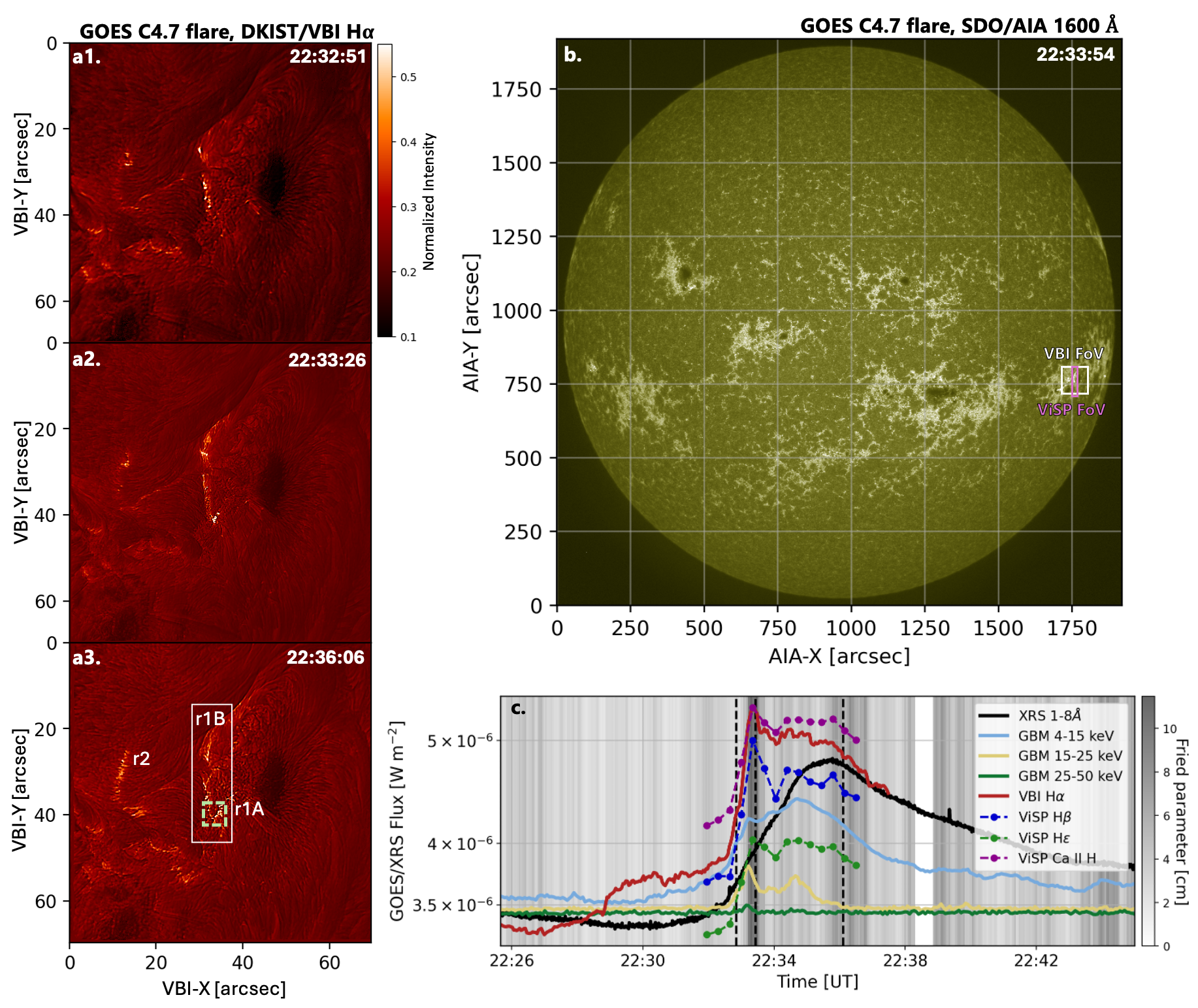}
    \end{center}
    \caption{Overview of the GOES C4.7-class flare on 11 August 2024. (a) DKIST/VBI H$\alpha$ images. The white box indicates the field of view of Figure \ref{fig:VBI_evolve}. (b) SDO/AIA 1600 \AA\ image at 22:33:54 UT.  The white and light violet boxes indicate the fields of view of VBI and ViSP, respectively.  (c) Light curves: GOES/XRS 1-8 \AA, \textsl{Fermi}/GBM in three spectral bands, VBI H$\alpha$ integrated intensity within the white box in panel (a), and the mean H$\beta$, H$\epsilon$, and Ca \textsc{II} H line intensities from 14 ViSP scans between 22:31:53 UT and 22:36:30 UT, including the entire ViSP field of view.  All except the GOES light curves are scaled and shifted along the y-axis for clarity.  DKIST light curves are limited to the period when the adaptive optics (AO) system was effectively locked.  Shading in panel (c) indicates the Fried parameter $r_0$.  Vertical dashed lines indicate the times of panels a1--a3.  In panels a1--a3 and Figure \ref{fig:VBI_evolve}, each frame is normalized independently so that intensity values lie between 0 and 1. An animation of the speckle-reconstructed VBI H$\alpha$ images for the region in panel (a) from 22:31:26 UT to 22:37:29 UT, including an inset showing the integrated H$\alpha$ light curve, is available online.}
    \label{fig:event_sum}
\end{figure*}

On 11 August 2024, a GOES C4.7-class flare occurred in active region 13777 at a heliocentric viewing angle of $\mu =$ 0.54. The flare start, peak, and end times were 22:29~UT, 22:35~UT, and 22:42~UT according to the GOES X-ray Sensor (GOES/XRS) 1--8 \AA\ channel light curve.  Figure \ref{fig:event_sum} shows the ribbon evolution in DKIST/VBI H$\alpha$ images, a 1600 \AA\ image from the Solar Dynamics Observatory Atmospheric Imaging Assembly (SDO/AIA; \citealt{lemen2011}) for context, and several flare light curves. 

We define three flare ribbon regions-of-interest within the VBI field of view: r1A, r1B, and r2 in Figure \ref{fig:event_sum}(a3).  We decompose the western flare ribbon r1 into two regions, r1A and r1B, as these develop at different stages and exhibit distinct spectral behavior (Section \ref{sec:kmeans}). As r1A and r1B evolve, there is a third brightening to the northwest that appears to be magnetically connected to AR~13777 based on images in the SDO/AIA 171, 193, and 1600 \AA\ channels (not shown).  

The Gamma-Ray Burst Monitor on the \textsl{Fermi} satellite \citep[\textsl{Fermi}/GBM;][]{2009ApJ...702..791M} collected hard X-ray (HXR) observations during the flare. Spatially resolved HXR spectra are unavailable.  In Figure \ref{fig:event_sum}(c) we show \textsl{Fermi}/GBM light curves from the sunward-facing detector NaI05. The \textsl{Fermi}/GBM spectra reveal two distinct periods of HXR emission at energies above 25 keV, from 22:32:40 UT to 22:33:20 UT (period 1) and 22:34:17 UT to 22:34:50 UT (period 2).

\subsection{DKIST observations}

DKIST observed the flare during the execution of Cycle 2 Experiment 2.11.\footnote{The DKIST data center currently maintains a few dozen potential flare datasets, listed at: \url{https://nso.atlassian.net/wiki/spaces/PDOI/pages/3600024522/DKIST+Flare+Catalog}. The data associated with this experiment are available at \url{https://dkist.data.nso.edu/?proposalId=pid_2_11}.} Observing conditions varied significantly.  The Fried parameter $r_0$ (a measure of the fluctuations in Earth's atmosphere, which impact the quality of ground-based observations) rose from $r_0 = 2-6$ cm during the pre-flare to $r_0>10$ cm during the flare impulsive phase (see Figures \ref{fig:event_sum}(c) and \ref{fig:slot_VBI}(b)).  During the decay phase, seeing fluctuated between poor ($r_0<5$ cm) and good ($r_0\approx6-9$ cm). VBI observed with the H$\alpha$ filter (centered at 656.3 nm with an effective FWHM of $\sim$0.65 \AA, 69\arcsec x 69\arcsec\ field of view, and spatial sampling 0\arcsec.017; DKIST data center product ID L1-HDGZO) and blue continuum filter (centered at 450.3 nm with filter width 4.1~\AA, 45\arcsec x 45\arcsec\ field of view, and spatial sampling 0\arcsec.011; DKIST data center product ID L1-CIURM).  

Even with favorable observing conditions, turbulence in the Earth's atmosphere degrades ground-based observations, introducing apparent motions in solar features not representative of solar dynamics.  To correct for this, the DKIST data processing pipeline produces a single speckle-reconstructed image from every 80 raw VBI images \citep{woger2008}.  The reconstructed VBI image sequence has a cadence of 2.66~s. To further correct for persistent atmosphere-induced fluctuations, we apply a destretching algorithm that stabilizes the image by shifting image subfields relative to a common reference using spatial cross-correlation.\footnote{Bitbucket repository: \url{https://bitbucket.org/dkist-community-code/destretch/src/main/}}

ViSP observed the Ca \textsc{II} H 396.8~nm (including H$\epsilon$ 397.0~nm; DKIST data center product ID L1-LFBAP) and H$\beta$ 486.1~nm (DKIST data center product ID L1-RYNAL) lines on arm 1 (75\arcsec\ slit length, 0\arcsec.0295 pix$^{-1}$ spatial sampling, 0.92 pm pix$^{-1}$ spectral sampling) and arm 2 (60\arcsec\ slit length, 0\arcsec.0236 pix$^{-1}$ spatial sampling, 1.1 pm pix$^{-1}$ spectral sampling) respectively.  The ViSP slit width was 0\arcsec.1. Using the methodology described in \cite{tamburri2026}, we conservatively estimate the spatial resolution of the ViSP to be as small as 0\arcsec.4-0\arcsec.5, though this varies considerably with seeing.

The DKIST/ViSP data were prepared by the DKIST data center team to level 1 (L1) using the standard reduction pipeline.\footnote{The coordinates provided in the VBI and ViSP L1 headers are incorrect.  In Figure 
\ref{fig:VBI_evolve} and throughout, we have approximately co-aligned ViSP and VBI helioprojective coordinates (HPC) using the coordinates reported in the H$\beta$ spectra as a reference.  We first align the spectra from the two arms of the ViSP using the hairlines visible in the intensity maps of both arms. We then align VBI with ViSP using common bright features visible in the VBI H$\alpha$ images and ViSP H$\epsilon$  intensity maps (similar to the method of \citealt{tamburri2026}). For our purposes we only require this approximate co-alignment between VBI and ViSP. This method carries the caveat that a feature visible in one chromospheric line may not be exactly co-spatial with the same feature observed in a different line. We note that the reported HPC coordinates in the L1 ViSP header appear offset from SDO by about $-$50'' and $-$40'' in x and y, respectively. } Each ViSP raster scan consisted of 91 slit steps, producing a dense $\sim$9\arcsec-wide raster.  The time between steps was 0.055~s for a scan time of 5~s, but the instrument overhead significantly lengthened the map cadence to 21 s. We calibrate the spectral axis and intensity using the Neckel solar atlas spectrum \citep{neckel1984,neckel1999} as described in \cite{tamburri2026}.  

The western flare ribbon (r1A and r1B) was entirely within the ViSP field of view.  In addition to these, VBI captured the eastern ribbon r2 and other small brightenings associated with the flare. The remote brightening visible to SDO is not within the DKIST field of view.

\section{Spatial structure of spectroscopic diversity in the chromosphere}\label{sec:ViSPvar}
\subsection{Evolution of flare ribbon in VBI images}

\begin{figure*}
\includegraphics[width=1\linewidth]{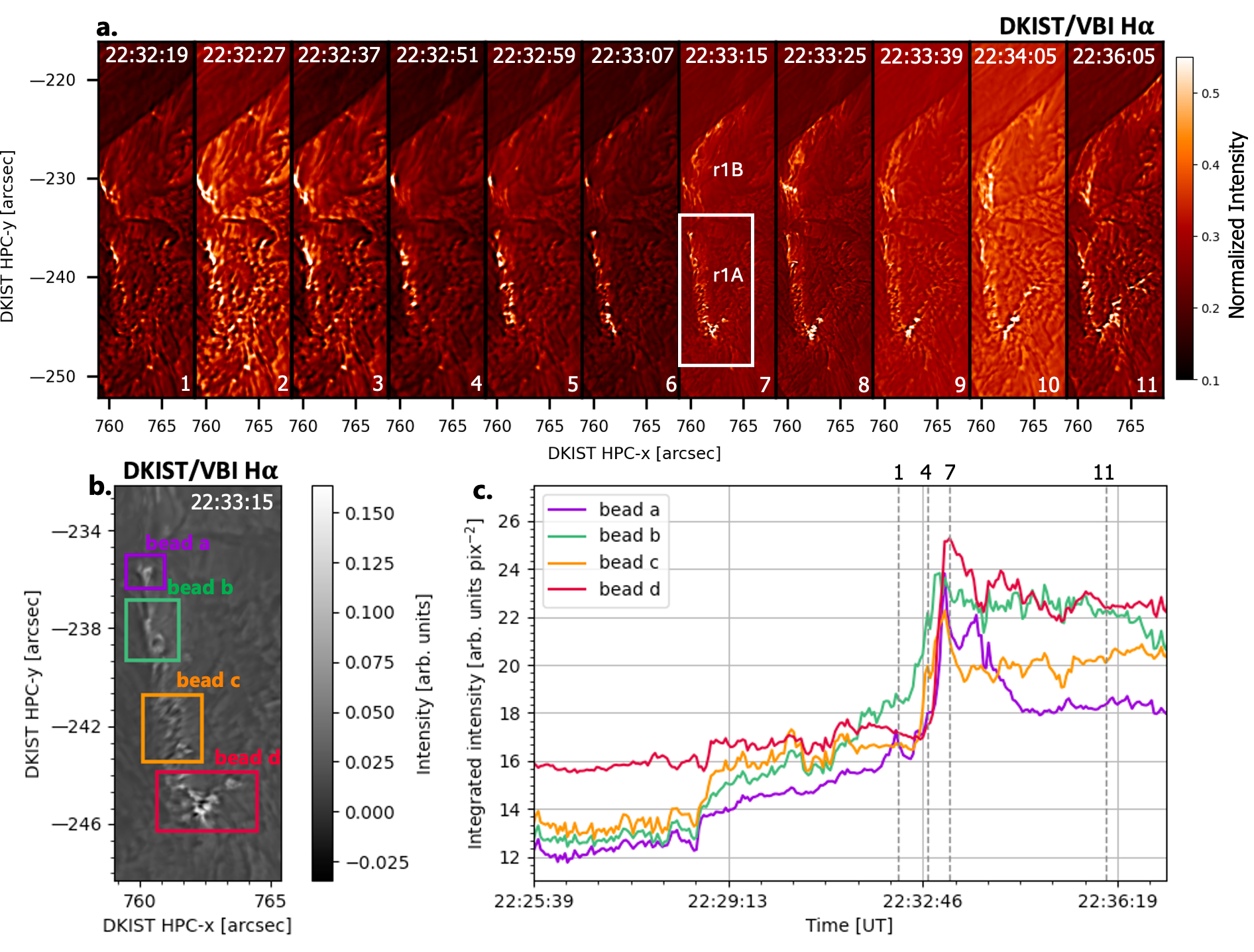}
    \caption{Evolution of the flare ribbon in VBI images. (a) Select VBI H$\alpha$ images of the flare ribbon within the white box in Figure \ref{fig:event_sum}(a3). (b) VBI H$\alpha$ image in the white box in (a), with colored boxes indicating the locations of beads visible in the ViSP spectra in Figure \ref{fig:kmeans_manyscan_caii}.  (c) H$\alpha$ light curves in the colored boxes in (b). Vertical lines indicate the times of four select numbered images in panel (a).}
    \label{fig:VBI_evolve}
\end{figure*}

Figure \ref{fig:VBI_evolve}(a) shows the evolution of the flare ribbons in the VBI H$\alpha$ filter.  Between $\sim$22:32:37 UT and 22:33:15 UT, the ribbon spreads southward along region r1A.  Region r1A then extends northward around 22:33:07 UT to produce a hook-shaped ribbon at helioprojective y-coordinate ($\text{HPC-y}$) $\approx-$245\arcsec.  Only a small brightening is evident in the northward region r1B during most of the evolution of r1A.  At roughly 22:33 UT, r1B begins to brighten more extensively. 

In Figure \ref{fig:VBI_evolve}(c) we show the integrated light curve for four areas within region r1A, corresponding to the locations of \textit{beads} evident in the ViSP spectra discussed below (e.g., Figure \ref{fig:ViSP_evolve}(c)). Although the beads are more obvious in the red wing of spectral lines observed by ViSP, for consistency we refer to these regions as beads in our discussion of the VBI images as well.

All bead light curves show a small increase in intensity around 22:28:30 UT followed by a gradual rise in the light curves for beads a, b, and c.  The gradual rise is possibly a chromospheric flare precursor, a phenomenon that has been studied in depth elsewhere (e.g., \citealt{panos2023,kumar2026,seyfritz_2026}). During the impulsive flare phase beginning around 22:32:30 UT, bead b brightens first, followed by beads a and c, and finally bead d, consistent with the generally southward expansion of the ribbon.  The peak of the light curve for bead a occurs later than that of bead b, indicating that the ribbon spreads slightly northward as well.

\subsection{Snapshots of the flare ribbon in ViSP spectra}

\begin{figure}
    \begin{center}\includegraphics[width=\linewidth]{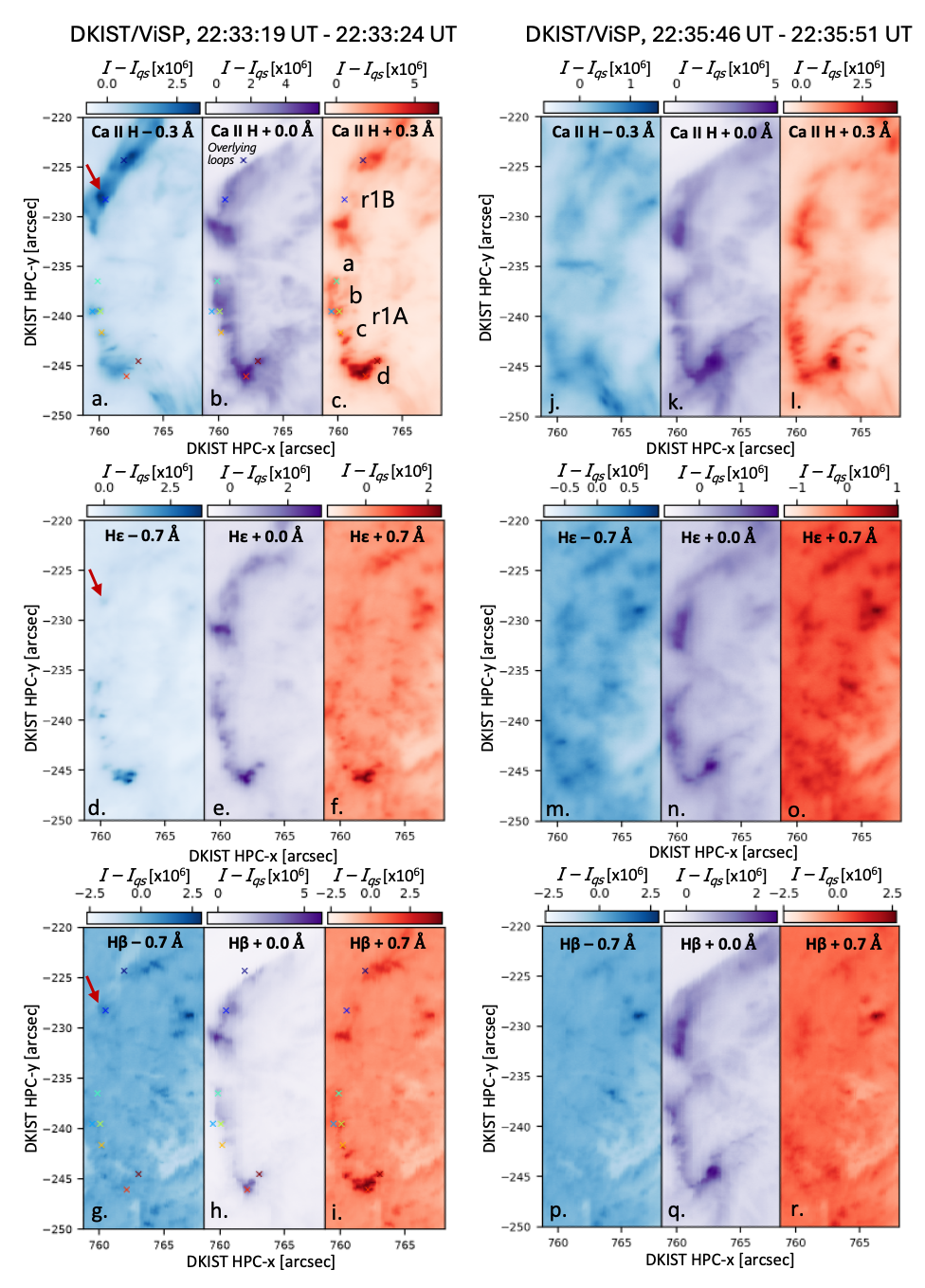}
    \end{center}
    \caption{Spectral structure of the flare chromosphere from ViSP within the white box in Figure \ref{fig:event_sum}(a). (a--c) ViSP Ca \textsc{II} H intensity in the blue wing, core, and red wing during the ViSP scan from 22:33:19 UT to 22:33:24 UT; (d--f) The same for H$\epsilon$. (g--i) The same for H$\beta$.  Panels (j--r) show the same as (a--i) during the ViSP scan from 22:35:46 to 22:35:51 UT, later in the flare. The spectra at the colored crosses in panels (a--c) and (g--i), selected by hand, are investigated in more detail in Appendix \ref{app1}. Point colors are not associated with the cluster colors in Figures \ref{fig:kmeans_manyscan_caii} and \ref{fig:kmeans_manyscan_hbeta}.}
    \label{fig:ViSP_evolve}
\end{figure}

The ViSP scan at 22:33:19 UT captured the flare impulsive phase and was collected under the best seeing conditions ($r_0 >$ 10 cm). In Figure \ref{fig:ViSP_evolve} we show the ribbon intensity in the blue wing, line core, and red wing of Ca \textsc{II} H, H$\epsilon$, and H$\beta$ for two ViSP scans: from 22:33:19 UT to 22:33:24 UT and approximately 2.5 minutes later, from 22:35:46 to 22:35:51 UT. Many of the same ribbon structures as in the H$\alpha$ images (Figure \ref{fig:VBI_evolve}) are evident in the ViSP intensity maps, including the overlying coronal loops obscuring part of r1B (see e.g., Figures \ref{fig:VBI_evolve}(a) and \ref{fig:ViSP_evolve}(b), $\text{y }>-$225\arcsec).  Prominent bead-like regions of red wing emission (labeled a--d) are visible in region r1A ($\text{y }<-$235\arcsec\ in all three chromospheric lines), especially during the impulsive phase scan (Figure \ref{fig:ViSP_evolve}(c, f, i)). We analyze these further in Section \ref{sec:kmeans}.

Both Ca \textsc{II} H and H$\epsilon$ have higher red wing than blue wing intensities in r1A.\footnote{Since the Ca \textsc{II} H and H$\epsilon$ are blended, the structure of the ribbon in the H$\epsilon$ blue wing is impacted by the red wing Ca \textsc{II} H intensity.  This produces bead-like structures in the H$\epsilon$ blue wing (Figure \ref{fig:ViSP_evolve}(d)) that are not present in the other lines.}  In much of r1B, there is elevated blue wing intensity in the Ca \textsc{II} H line.  There is a compact, blueshifted core at roughly [760\arcsec,$-$228\arcsec] (red arrow in panels (a), (d), and (g)) visible in the blue wing of all three lines during the first ViSP scan from 22:33:19 to 22:33:24 UT. Whereas the rest of the blueshifted region is smooth, spatially extended, and relatively transient (especially in the Ca \textsc{II} H blue wing), the redshifted region of the ribbon demonstrates long-lasting bead-like structures.

The smooth, extended morphology of the r1B blueshifts in contrast to the mostly redshifted, fragmented r1A is notable. In the H$\beta$ blue wing (Figure \ref{fig:ViSP_evolve}(g)) much of r1B is slightly in absorption.  Additionally, the SDO/AIA 1600 \AA\ images (not shown) reveal a remote brightening that appears concurrently with the brightening of r1B.  Region r1B appears magnetically connected to this remote brightening based on SDO/AIA 193 and 131 \AA\ images (not shown). It follows that the blueshifted feature could be a jet from the reconnection region in a ``fan-spine''-type reconnection scenario (e.g., \citealt{duan_2024_jet}).  

The compact brightening in r1B located between roughly $\text{y }=-$227\arcsec\ and $\text{y }=-$232\arcsec\ is still present three minutes into the decay phase, although the blueshifted component has largely disappeared, leaving only an unshifted or slightly redshifted component (e.g., Figure \ref{fig:ViSP_evolve}(j--l)).  

\subsection{\textit{k}-means clustering of ViSP spectral profiles}\label{sec:kmeans}

The chromospheric spectra are remarkably diverse.  To characterize the dominant spectral features in the chromosphere and investigate their spatial distribution, we employ the \textit{k}-means clustering technique (e.g., \citealt{bird2009}). There are several previous studies of flare spectra that use a similar methodology (e.g., \citealt{panos2018,french2025_downflows,yadav2025}). 

\textit{k}-means is an unsupervised machine learning algorithm that assumes each member of a dataset is described by a predefined number of features $n$ (in this case, the features are the normalized intensity values $I_{\lambda}^{norm}$ at each wavelength $\lambda$ from the far blue wing to the far red wing of each spectral line considered). The algorithm first initializes $k$ cluster centers within an $n$-dimensional space (either randomly or, when using \texttt{kmeans++} initialization, by probabilistically selecting initial cluster centers to be farther from already-chosen centers), where $k$ is determined by the user.  The algorithm then measures the Euclidean distance of each member of the dataset to each cluster center and groups the members according to the nearest cluster center. Cluster centers are then recalculated using the mean of all members of each cluster.  The process iterates until the distances to the cluster centers are stable or a maximum set of iterations is reached.

For our implementation we use the \texttt{cluster.KMeans} method within the \texttt{scikit-learn} Python package.\footnote{\url{https://scikit-learn.org/stable/modules/generated/sklearn.cluster.KMeans.html}}  We generally use the default Python parameters, including the \texttt{kmeans++} initialization to speed convergence, except that we increase the number of initializations with different centroid seeds from the default ($n_{init}=1$) to $n_{init}=100$ to improve the accuracy of the clustering.

Before clustering, we subtract an estimated non-flare spectrum from the flare spectra. This will expedite future model-data comparisons, when it is often necessary to isolate the flare-associated spectral signal from the background. We estimate the non-flare spectrum by averaging the spectra in space and time from the scan between 22:33:19 UT and 22:33:24 UT in a location where there is no clear influence from solar features present in the flaring portion of the active region.\footnote{Ideally, we would remove the pre-flare from each pixel independently using an earlier ViSP scan.  However, seeing conditions were suboptimal during the scans preceding the flare (see Figure \ref{fig:event_sum}(c)). The resulting pointing drift of the telescope means that the ViSP field of view shifts slightly between the pre-flare and the flare-time scans.  The best option, therefore, is to estimate the non-flare spectrum using a portion of the active region that does not participate in the flare, with spectra from the best-seeing scan.  We acknowledge that this choice may have a slight impact on our results especially given the optically-thick nature of the lines studied, though most of the flare emission should overwhelm impacts of the non-flare spectrum.} 

We then generate separate ViSP intensity maps for Ca \textsc{II} H and H$\beta$ using the mean intensity between the blue and red line wings in each ViSP pixel. We produce a flare ribbon mask for each map by selecting only the pixels in the scan for which the pixel intensity exceeds 1.5 $\times\; I_{med}$ and 3 $\times \;I_{med}$ for Ca \textsc{II} H and H$\beta$ respectively, where $I_{med}$ is the median intensity in all scans for the line in question after subtracting the non-flare spectrum. The intensity cutoffs for ribbon masking are fairly liberal in order to capture as much of the flare ribbon as possible.  As a result, especially in later scans or during poor seeing, there may be some non-flare pixels included in the mask.  The masking is more challenging for the H$\beta$ spectra, when spurious bright pixels unrelated to the flare are often included in the mask (e.g., the second panel in Figure \ref{fig:kmeans_manyscan_hbeta}(a)). This does not affect the  primary conclusions of the $k$-means analysis, although it likely produces some anomalous H$\beta$ clusters.

Our primary interest lies in the dynamics revealed by spectral line shape rather than line intensity. We therefore normalize the spectral profiles so that the intensities lie between 0 and 1 to ensure that variations in absolute intensity do not affect the clustering results. We then cluster all masked spectra from ten ViSP scans beginning at 22:32:58 UT simultaneously to track both spatial and spectral evolution throughout the flare. We perform the $k$-means analysis on the Ca \textsc{II} H and H$\beta$ spectra separately and determine based on three performance metrics and visual inspection that $k=$ 12 and 14 are the optimal number of clusters for Ca \textsc{II} H and H$\beta$ respectively. We present a justification for this result in Appendix \ref{app0}. 

Figures \ref{fig:kmeans_manyscan_caii} and \ref{fig:kmeans_manyscan_hbeta} show the clustering results for Ca \textsc{II} H and H$\beta$, respectively. We order the representative profiles (RPs) in Figures \ref{fig:kmeans_manyscan_caii}(b) and \ref{fig:kmeans_manyscan_hbeta}(b) according to the intensity-weighted mean wavelength, where weights are the intensity values at each wavelength position.

We quantify the velocity shift of the cluster center profiles using the 30\% bisector, the midpoint of the spectral positions at which the red and blue line intensities are 30\% of the peak line intensity.   This is a standard method for approximating signatures of bulk flows (e.g., condensations) in optically thick lines \citep[][and references therein]{graham_cauzzi_2015}. For the Ca \textsc{II} H clustering, the 30\% bisector of cluster 1 (the most blueshifted), converted to a velocity shift, is $v = -11$ km s$^{-1}$; the 30\% bisector of cluster 12 (the most redshifted) corresponds to a velocity shift of $v = 22$ km s$^{-1}$. For the H$\beta$ clustering, the 30\% bisector of cluster 1, converted to a velocity shift, is $v = 1.2$ km s$^{-1}$; the 30\% bisector of cluster 13 corresponds to a velocity shift of $v = 21$ km s$^{-1}$.  The peak of RP1 in the H$\beta$ clustering is blueshifted with a velocity of $v\approx-$20 km s$^{-1}$. The blueshift is less prominent in the H$\beta$ clustering results than for Ca \textsc{II} H.  This may be because the lines are sensitive to distinct physical processes. For example, (i) hydrogen Balmer series lines are predominantly sensitive to electron density \citep[e.g.,][]{kowalski2022,tamburri2026}, such that other factors contributing to the width of H$\beta$ are less prominent; (ii) during a flare, Ca \textsc{ii} lines may form over a broader range of heights than H$\beta$ \citep{kuridze_2020} and may therefore be subject to a wider range of plasma processes; and (iii) opacity differences between the two lines in the flaring chromosphere can produce different line shapes.  Further investigation of these differences is warranted.

\begin{figure*}
     \begin{center}
    \includegraphics[width=.85\linewidth]{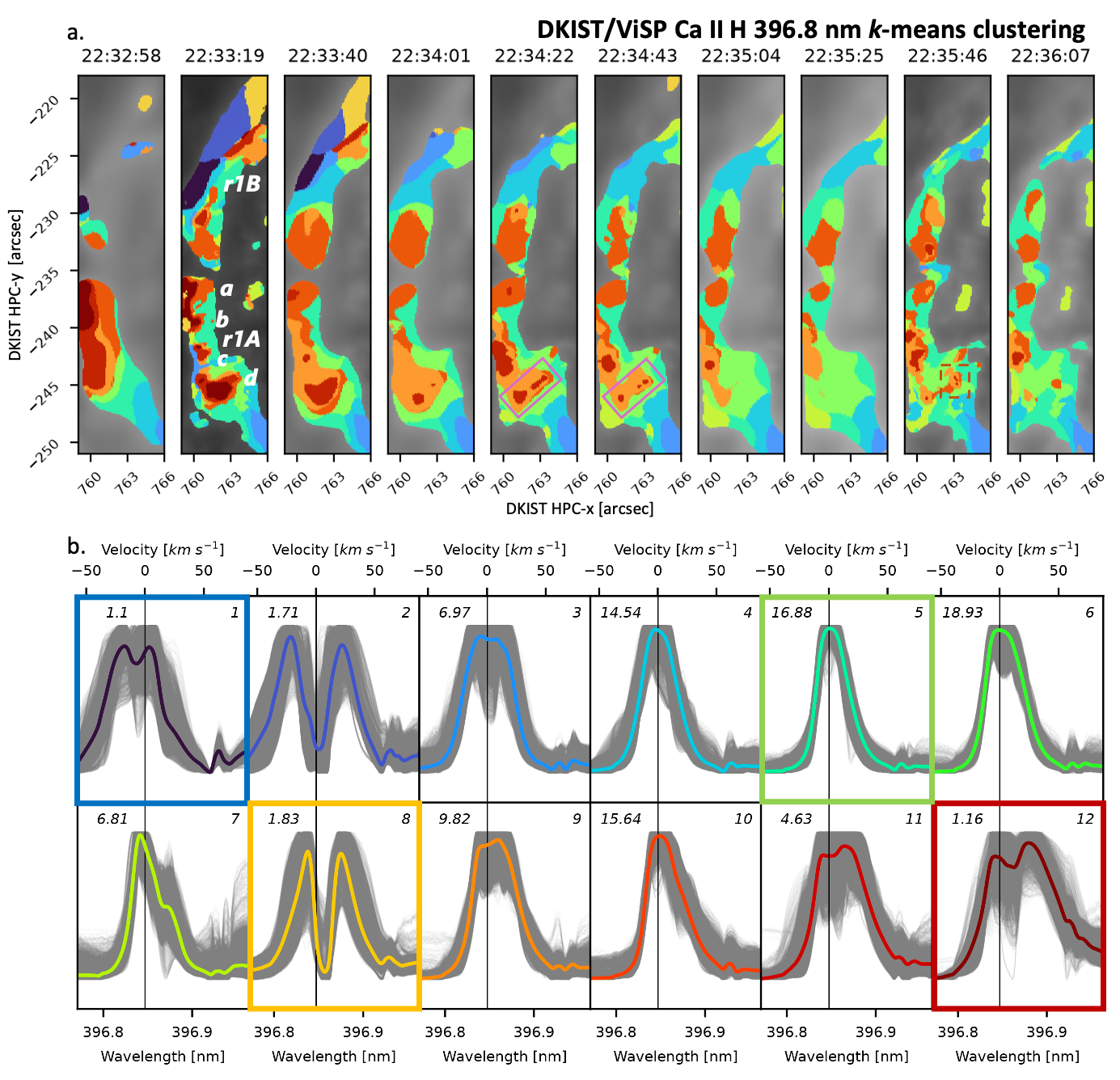}
    \end{center}
    \caption{\textit{k}-means clustering of ViSP Ca \textsc{II} H spectra during the 11 August 2024 flare. (a) ViSP mean intensity of the Ca \textsc{II} H line with the overlaid ribbon mask in color. Ribbon beads a--d and regions r1A and r1B are noted in the second scan.  Colors correspond to the cluster centers resulting from the \textit{k}-means clustering method shown in (b). In (b) the Ca \textsc{II} H profiles associated with each cluster are shown in gray.  RPs for each cluster are in color and ordered according to weighted mean of wavelength.  Blue, green, yellow, and red boxes around select clusters indicate RPs clearly exhibiting the four general spectral types noted in this work: blueshifted, symmetric, centrally-reversed, and redshifted spectral lines. Values in the upper left of each panel indicate the percentage of total spectral profiles contained in that cluster.}
    \label{fig:kmeans_manyscan_caii}
\end{figure*}

\begin{figure*}
     \begin{center}
    \includegraphics[width=.85\linewidth]{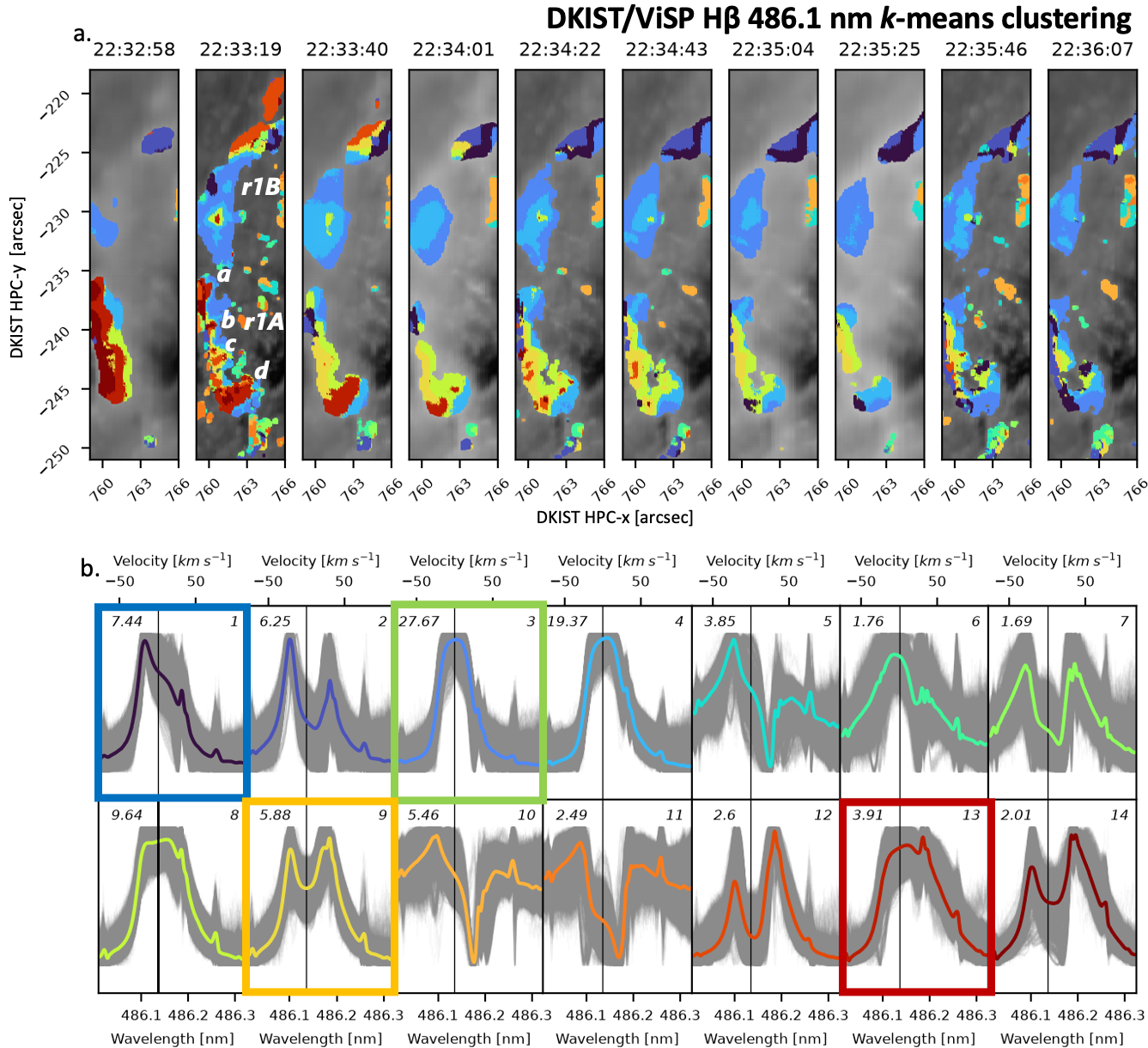}
    \end{center}
    \caption{The same as Figure \ref{fig:kmeans_manyscan_caii}, but for H$\beta$.}
    \label{fig:kmeans_manyscan_hbeta}
\end{figure*}

\subsection{Evolution of spectral clusters}

The \textit{k}-means analysis clearly characterizes the temporal evolution of pixel spectra. The spectra in r1A during the scan from 22:33:19 to 22:33:24 UT reveal a fragmented flare ribbon, notably the four \textit{beads} of centralized flare emission noted above (a, b, c, and d in Figures \ref{fig:kmeans_manyscan_caii}(a) and \ref{fig:kmeans_manyscan_hbeta}(a)). Beads a, b, and c are initially heated during or slightly prior to the ViSP scan at 22:32:58 UT, while bead d is initially heated after this scan and before the following scan at 22:33:19 UT. 

During the scan from 22:33:19 UT to 22:33:24 UT, central pixels in each bead are primarily redshifted (clusters 12 and 13 in Figures \ref{fig:kmeans_manyscan_caii}(b) and \ref{fig:kmeans_manyscan_hbeta}(b) respectively).  The pixels in the bead periphery are dominated by symmetric, single-peaked profiles (e.g., profiles 5 and 3 in Figures \ref{fig:kmeans_manyscan_caii}(b) and \ref{fig:kmeans_manyscan_hbeta}(b) respectively) or central reversals (e.g., clusters 8 and 9 in Figures \ref{fig:kmeans_manyscan_caii}(b) and \ref{fig:kmeans_manyscan_hbeta}(b) respectively).  Spectra in bead c (which may itself be composed of two or more heating events; see also Figure \ref{fig:ViSP_evolve}(c) and (f)) are mostly redshifted or symmetric.  Region r1B is dominated by blueshifted Ca \textsc{II} H profiles (cluster 1 in Figure  \ref{fig:kmeans_manyscan_caii}(b)). 

The full evolution of bead d, where flare emission is strongest (Figure \ref{fig:ViSP_evolve}(c), (f), and (i)), is well-captured by the ViSP.  At first, during the scan beginning 22:33:19 UT, the spectra are redshifted in the center of the bead. The periphery of the bead is generally unshifted.  Between 22:33:40 UT and 22:35:04 UT, the redshifts in bead d become less prominent.  During the scan at 22:33:19 UT, Ca \textsc{II} H spectra from most of this region are clustered with RP 12 in Figure \ref{fig:kmeans_manyscan_caii}(b), but at later times RP 11 (e.g., the scan at 22:33:40 UT) and RP 9 (e.g., the scan at 22:34:01 UT) are more prevalent. This evolution is in part due to degraded seeing conditions between 22:33:40 UT and 22:35:25 UT.  The seeing is again favorable starting at 22:35:46 through 22:36:07 UT.  At this point the spectra are less redshifted than in earlier scans. While most of bead d is unshifted and symmetric at this stage (e.g., RP 5), many pixels in bead c show small redshifts.  A region of bead d remains slightly redshifted even in the final clustered scan. Since the \textit{Fermi}/GBM 15-25 keV light curve shows a second period of enhancement around this time, the redshifts during this scan may indicate continued reconnection and flare heating via electron acceleration.

The blueshifts in region r1B are generally transient, only visible during the scan at 22:33:19 UT.  The southernmost ``core'' portion of r1B (directly east and north of a redshifted region at y $=-$230\arcsec) remains in emission throughout the ten scans, although the strongest blueshifts are only visible at 22:33:19 and 22:33:40 UT. By the final scan shown in the clustering results, r1B is mostly unshifted, with a small area containing slight redshifts.  Most of the same dynamics discussed here are evident in the H$\beta$ clustering as well, although line shifts sometimes differ.

\subsection{Summary of clustering results}

The \textit{k}-means analysis of Ca \textsc{II} H and H$\beta$ spectra reveals notably diverse spectra distributed in space and time, including: (i) broad, significantly \textit{redshifted} pixels; (ii) \textit{symmetric, single-peaked} profiles; (iii) \textit{central reversals}; and (iv) \textit{blueshifted} pixels. The select representative profiles indicated by colored boxes in Figures \ref{fig:kmeans_manyscan_caii}(b) and \ref{fig:kmeans_manyscan_hbeta}(b) are examples of these four main categories, though there is variation in the exact spectral shape within each category. Some \textit{anomalous} cluster centers (e.g., RP 5 in Figure \ref{fig:kmeans_manyscan_hbeta}(b)) are also present.  In many cases, these are due to the inclusion of bright non-flare pixels in the ribbon mask.

While the symmetric profiles are very common (e.g. 28\% of H$\beta$ profiles are clustered with RP 3 alone), the other three types are not.  This is in part because the extreme or unique conditions necessary to produce strong blueshifts, redshifts, or central reversals are often transient or uncommon, while symmetric, unshifted emission lines are often produced by more long-lasting gradual processes \citep[e.g.,][]{tamburri2026}.

The clustering method effectively captures salient aspects of the spectral dynamics observed by ViSP. In Appendix \ref{app1} we compare the evolution in select pixels of the ViSP to corresponding $k$-means clusters to illustrate this point.

\section{Ribbon substructure in VBI images and ViSP spectra}\label{sec:finestructure}

\begin{figure*}
     \begin{center}
    \includegraphics[width=.95\linewidth]{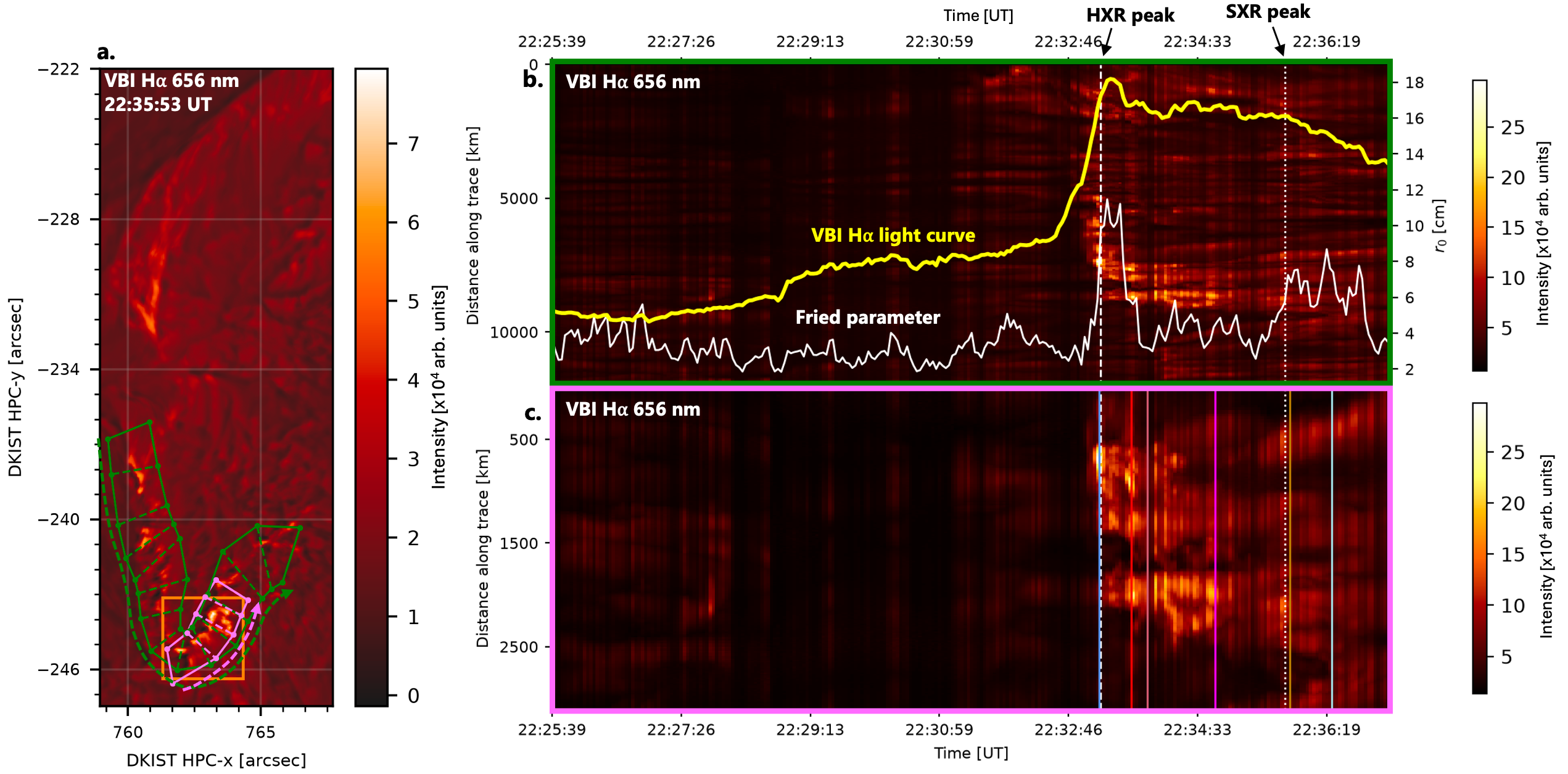}
    \end{center}
    \caption{Evolution of flare ribbon region r1A. (a) VBI H$\alpha$ image at 22:35:53 UT, showing green and magenta slots. Green and magenta dotted arrows indicate the direction of increasing distance along the trace. The dark orange box indicates the field of view of Figure \ref{fig:VBI_ultrafine}.  (b) Time-distance plot of ribbon trace intensity within the green slot in panel (a). The VBI H$\alpha$ light curve, Fried parameter, and times of the peak HXR and SXR flux are shown in yellow, solid white, dashed white, and dotted white lines respectively.  (c) Time-distance plot of ribbon trace intensity within the magenta slot in (a).}
    \label{fig:slot_VBI}
\end{figure*}

In this section we analyze the dynamic, multiscale nature of ribbon substructure revealed by VBI.  We first investigate the evolution of ribbon fine structure in region r1A (Figure \ref{fig:slot_VBI}).  We detail the apparent splintering of ribbon beads into smaller-scale features as the flare proceeds.  We show that both $\sim$300--470 km wide \textit{bead-like} structures and $\sim$50 km wide \textit{loop-like} structures are visible at different times within the same region of the flare ribbon (Figure \ref{fig:VBI_ultrafine}), but note that this apparent splintering may be largely due to variations in observing conditions.  We show that the smallest observable features are co-spatial with Ca \textsc{II} H redshifts, thereby connecting the observed spatial fine structure with dynamics inferred from high-resolution spectra (Figure \ref{fig:ViSP_ultrafine}).

\subsection{Determination of the ribbon trace}\label{sec:trace}

To track the evolution of small-scale ribbon features, we employ the following semi-autonomous method: For all frames, we manually define a common \textit{slot} that bounds the region within which the ribbon evolves (green and magenta bounds in Figure \ref{fig:slot_VBI}(a)). We then define the unique 1D ribbon \textit{trace} for each frame by tracking the position and magnitude of the maximum intensity along perpendicular lines at each point along the ribbon.  This accounts for the apparent motion of the ribbon and successfully identifies variations in bead-like structures along the ribbon trace. In Figure \ref{fig:slot_VBI} we show time-distance plots of ribbon intensity for the entire region r1A (panel b) and an isolated portion of the ribbon (panel c) from 22:25 UT to 22:37 UT.  

\subsection{Apparent splintering of flare ribbon beads}\label{sec:splintering}

Ribbon bead d, identified previously in the clustering of ViSP spectra (e.g., Figure \ref{fig:kmeans_manyscan_caii}(a)), is evidently composed of several smaller structures in the VBI H$\alpha$ images.  Several of these structures are visible in the lower, hook-shaped portion of region r1A (Figure \ref{fig:slot_VBI}(c)). The two northernmost beads (at roughly 1500 and 2000 km in Figure \ref{fig:slot_VBI}(c)) appear to splinter into even smaller structures around 22:35 UT.  However, some of this apparent splintering occurs concurrently with improved observing conditions around 22:35:30 UT, suggesting that the resolution of smaller structures is largely due to variations in observing conditions.  In Appendix \ref{sec:crosssections} we present further evidence to this end. We conclude that the appearance of ribbon fragments may be strongly related to observing conditions, although the larger ribbon beads may still have physical significance, as discussed further in Section \ref{sec:disc_c6}.

\begin{figure*}
     \begin{center}
    \includegraphics[width=1\linewidth]{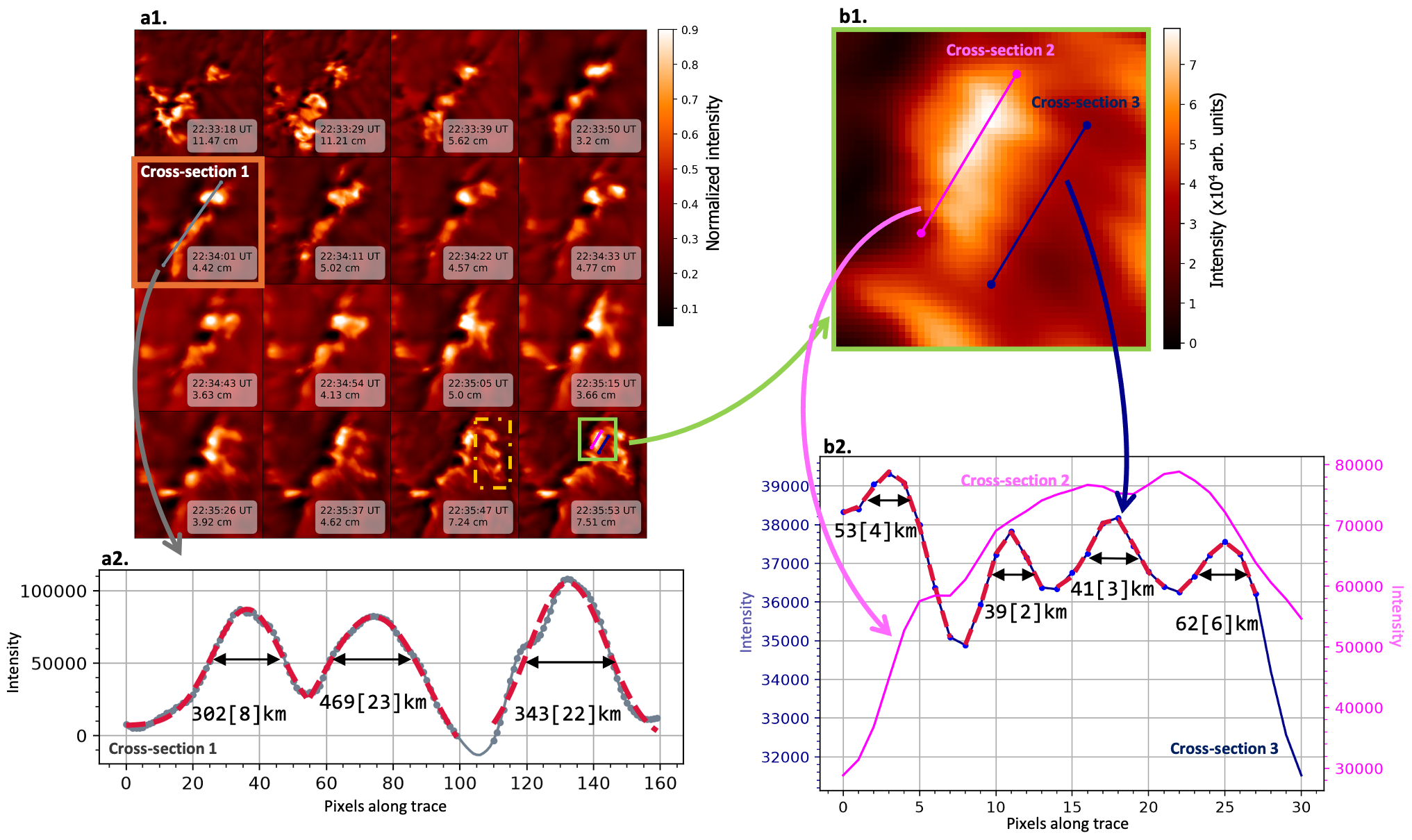}
    \end{center}
    \caption{Analysis of select flare ribbon structures in the southern portion of region r1A. (a1) Time evolution of the orange-outlined region in Figure \ref{fig:slot_VBI}(a) from 22:33:18 UT to 22:35:53 UT. The grey line intersects the three beads visible in this frame (cross-section 1).  In each panel we note the corresponding time and Fried parameter.  (a2) Intensity along cross-section 1, with widths and standard deviation errors  of ribbon structures shown. (b1) Ribbon bead in the light-green outlined region in (a1), showing two cross sections: cross-section 2 (magenta) across the brightest portion of the bead, and cross-section 3 (dark blue) perpendicular to loop-like structures near the bead. (b2) Intensities along cross-sections 2 and 3, with the widths and standard deviation errors of the loop-like structures in cross-section 3 indicated. An animation of the speckle-reconstructed VBI H$\alpha$ images for the region in panel (a) from 22:31:26 UT to 22:37:29 UT, including an inset showing the integrated H$\alpha$ light curve, is available online.}
    \label{fig:VBI_ultrafine}
\end{figure*}

Measurements of the smaller ribbon beads visible in the VBI H$\alpha$ image elucidate the apparent multiscale nature of the ribbon.  In the orange-outlined region in Figure \ref{fig:VBI_ultrafine}(a1), there are three bead-like structures visible especially between 22:33:40 and 22:36:20 UT.  We measure the widths of these structures to be between 300 and 470 km (Figure \ref{fig:VBI_ultrafine}(a2)).\footnote{Here and throughout, feature width is defined as the full width at half maximum (FWHM) of a Gaussian fit to the intensity profile across each feature, including a linear background to account for the background intensity. The fitting is done with \texttt{scipy.optimize.curve\_fit}.}  By roughly 22:35:30 UT, the southernmost beads are no longer coherent.  The northernmost bead, which is the brightest throughout the evolution of the region, eventually splinters into loop-like structures 39--62 km wide (Figure \ref{fig:VBI_ultrafine}(b1--b2)).

\subsection{Spectroscopic properties of ribbon fragments}\label{sec:splintspec}

While the ViSP spectra (with spatial sampling of 0\arcsec.03 along the slit in arm 1 and 0\arcsec.1 in the scanning direction) are inherently subject to more restrictive resolution limits than the VBI (0\arcsec.017 spatial sampling in H$\alpha$), bead-like structures are still visible.  For example, during ViSP scans at 22:34:22 UT and 22:34:43 UT, the Ca \textsc{II} H \textit{k}-means clustering results clearly identify the three bright regions visible in r1A in the VBI images (e.g., Figure \ref{fig:VBI_ultrafine}(a1)).  The pixels in these three locations (outlined by two magenta boxes in Figure \ref{fig:kmeans_manyscan_caii}(a)) are clustered in RPs 11 and 9, which both have a slightly elevated red wing in Ca \textsc{II} H.

Later, during the ViSP scan at 22:35:46 UT, the bead that appears to fragment into smaller loop-like structures in the VBI images (Figure \ref{fig:VBI_ultrafine}) also exhibits distinct spectral behavior (orange dashed box in Figure \ref{fig:kmeans_manyscan_caii}(a), an isolated region of pixels clustered into RP 11). In this section we further investigate the spectroscopic behavior of this feature.

\begin{figure*}
     \begin{center}
    \includegraphics[width=1\linewidth]{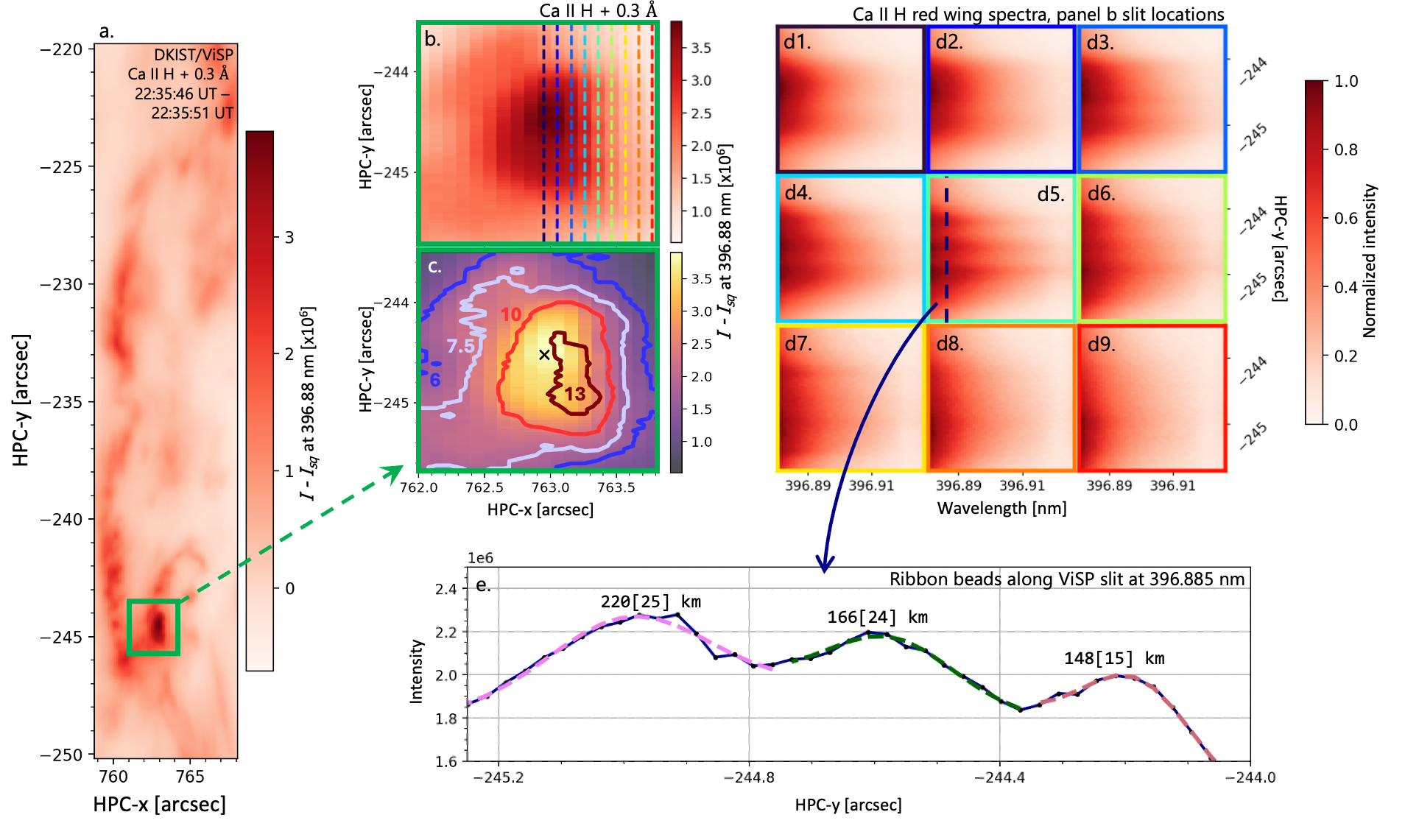}
    \end{center}
    \caption{Fine structure in chromospheric ViSP spectra.  (a) Intensity in the red wing of Ca \textsc{II} H during the scan at 22:35:46 UT.  (b) The northernmost bead in Figure \ref{fig:VBI_ultrafine}(a1). (c) Intensity of Ca \textsc{II} H at 396.88 nm for the bead in (b), with contours showing the 30\% bisector for the spectra within this frame.  Numerical values for the shown contours are given in km s$^{-1}$. The black `x' indicates the pixel of peak intensity in the Ca \textsc{II} H line wing during this scan. (d1--d9) Ca \textsc{II} H red wing spectral intensity at the ViSP slit positions indicated by the vertical lines in (b), normalized for each slit position to clearly show spatial structures for all locations.  (e) Intensity along the black vertical line in (d5), intersecting the structures observed in the Ca \textsc{II} H red wing, with reported widths and errors.}
    \label{fig:ViSP_ultrafine}
\end{figure*}

The spatial sampling is better along the ViSP slit than in the scanning direction. Fortunately, the ViSP slit intersects the small-scale structures observed in the VBI images at a nearly-perpendicular angle.  We can leverage this orientation to analyze the spectral characteristics of the ribbon at this location.   During the ViSP scan at 22:35:46 UT (Figure \ref{fig:ViSP_ultrafine}), the bead-like structures are characterized by slight intensity enhancements in the red wing of Ca \textsc{II} H between 396.87 nm and 396.92 nm.  

To quantify the associated redshift in this bead, we calculate the 30\% bisector of Ca \textsc{II} H in each pixel.  The contour map of this quantity in Figure \ref{fig:ViSP_ultrafine}(c) shows that the line shift peaks at just over 13 km s$^{-1}$.  The peak velocity is offset from the location of peak red wing Ca \textsc{II} H intensity (black `x', panel c) in the direction of the loop-like features observed in the VBI images (Figure \ref{fig:VBI_ultrafine}), indicating that the strongest downflows occur in these strands rather than in the bead center.

Ca \textsc{II} H spectra at the raster position indicated in Figure \ref{fig:ViSP_ultrafine}(b) reveal 148--220 km wide structures along the slit (panel e).  We suggest that the ViSP slit likely crosses the three vertical structures outlined by a dashed orange box in the second to last panel of Figure \ref{fig:VBI_ultrafine}(a1) at 22:35:47 UT, one of which later splinters further into the small-scale loop-like structures shown in Figure \ref{fig:VBI_ultrafine}(b1).  

The ViSP is unable to resolve the smallest loops observed by VBI. However, since the bead fragments are characterized by elevations in the red wing intensity of Ca \textsc{II} H, it is reasonable to conclude that the ViSP and VBI have observed downflowing material along magnetic loops in this region.  

\section{Discussion}\label{sec:disc_c6}

Using DKIST observations of a C-class flare, we have presented two important manifestations of flare fine structure: (i) the spectral diversity of the chromospheric Ca \textsc{II} H and H$\beta$ lines when observed at high resolution, and (ii) the multiscale nature of ribbon substructure.  

\subsection{Implications of spectroscopic diversity in the flare ribbon}

The flare-time Ca \textsc{II} H 396.8 nm and H$\beta$ 486.1 nm spectra are primarily redshifted, unshifted and symmetric, reversed in the line core, or blueshifted, with significant variations over relatively small spatial scales (Figures \ref{fig:kmeans_manyscan_caii} and \ref{fig:kmeans_manyscan_hbeta}). This analysis complements previous studies by \cite{panos2018}, \cite{pietrow2024}, \cite{faber2026}, \cite{kerr2026}, and others in demonstrating that flare spectral profiles are particularly varied when observed at high spatial resolution.

The variety of chromospheric spectra observed in a single flare is important when modeling a flare atmosphere using modern radiative-hydrodynamic codes such as RADYN \citep{carlsson1992,carlsson1997,allred2005,allred2015,allred2020}. RADYN has been used extensively to forward-model the atmospheric response to heating by comparing synthetic spectra to high-resolution spectra from several observatories. Such efforts have successfully reproduced many salient and distinct features of the high-resolution spectra, including a broad range of spectral line characteristics and atmospheric dynamics (e.g., \citealt{polito_2018,kerr_2019_siIV,polito_2023,sadykov2023,kerr_2024,lorincik_2025,kerr2022r,kerr2023r,kerr2026}) and the broadening of the prominent Balmer lines of hydrogen (e.g., \citealt{kowalski2017b,kowalski2026}), even in DKIST flare data \citep{tamburri2026}. 

However, given the lack of unequivocal observational constraints for many flare heating scenarios, it remains challenging to self-consistently explain the entirety of any one flare's demonstrated spectral behavior.  For example, the physical parameters describing a theorized beam of non-thermal electrons injected into the chromosphere can be broadly derived from the hard X-ray spectra collected by \textit{Fermi}/GBM or the Spectrometer/Telescope for Imaging X-rays (STIX) onboard Solar Orbiter (although uncertainties still exist in the derived beam parameters such as the low-energy cutoff and flux density). A similar observational diagnostic does not yet reliably exist for beams of non-thermal protons except for certain $\gamma$-ray lines produced by collisions between accelerated ions above $\sim$1 MeV (including protons and $\alpha$ particles) with ambient chromospheric plasma \citep[e.g.,][]{aschwanden2002,vilmer2011}. Additionally, when relying on forward modeling to discern which heating mechanism is likely active, different heating scenarios can produce degenerate spectral signatures \citep[e.g.,][]{lorincik_2025,tamburri2026}. As the spatial resolution of solar observations has improved and the breadth of modeled flare heating scenarios has widened, both parameter spaces -- the variety of spectral signatures and the range of potential sources of flare heating -- have increased in range while our ability to connect the two has not improved accordingly.

Further investigations detailing the specific high-resolution spectral signatures that each heating scenario produces are therefore essential. The observations presented here reveal the broad range of spectral shapes within flaring sources. That they vary on such small spatial scales is likely the result of variations in flare processes. By resolving this spectral diversity using modern observations, we can probe flare energy transport on substantially smaller scales than afforded by direct signatures such as X-rays or $\gamma$-rays. 

Given this flare-time spectral complexity, the \textit{k}-means clustering technique (Section \ref{sec:kmeans}) is a powerful tool that can be used for rigorous identification of key spectral features. The method identifies the primary spectral dynamics that forward-modeling codes like RADYN need to reproduce, while also illuminating the spatial distribution of spectral types within the flare ribbon (e.g., Figure \ref{fig:kmeans_manyscan_caii}(a)).  It is notable that, although we normalized for pixel intensity prior to the $k$-means analysis and did not include spatial information in the clustering, the identified spectral clusters are clearly spatially coherent, especially in the Ca \textsc{II} H line (Figure \ref{fig:kmeans_manyscan_caii}), indicative of spatial structuring of flare heating scenarios.

By forward-modeling the key spectral signatures identified by the machine learning technique, we can resolve remaining degeneracies producing these features. To give a simple example, several scenarios can generate the central reversals observed in the DKIST chromospheric spectra. Reversals can be produced by line-specific radiative transfer effects such as the height-dependent source function for an optically thick line (e.g., \citealt{kuridze_2015b,kerr_2016}).  Alternatively, strong wing emission sometimes resembling a central reversal can be generated by deep chromospheric heating, for example in an Ellerman bomb scenario (e.g., \citealt{ellerman1917,hong2017,reid2017}), or even \textit{self-absorption} from optically thick overlying plasma. By modeling the response of the solar atmosphere to different particle populations that penetrate to different atmospheric depths, RADYN can be used to separately simulate many of these potentially relevant scenarios.  A detailed comparison of the synthetic and observed spectra may then provide evidence for the most likely active heating mechanisms in any given pixel. By scaling this method, important new constraints can be placed on the spatial distribution of active heating scenarios -- and their corresponding spectral signatures -- in solar flares.

\subsection{Implications of multiscale fine structure in the flare ribbon}

Our second primary finding is that VBI images and ViSP spectra reveal ribbon substructure across a wide range of spatial scales.  Observations of such multiscale ribbon fragmentation have been limited by the capabilities of most modern telescopes and the very few flares yet observed by DKIST. In the past, \cite{faber2025}, \cite{yadav2025}, \cite{french2025_beads}, \cite{collier_2026}, and others have noted ribbon substructure (and even periodicity) at scales of 150--500 km.  While the physical cause of these features is unclear, fine structure in flare ribbons is evidently common and even reflected in some ribbon proxies synthesized with 2.5D or 3D MHD flare simulations (e.g., \citealt{li2025,dahlin_2025}).  At the same time, it has been shown that the VBI can resolve coronal loop structures as small as 21 km wide, averaging 45--55 km \citep{tamburri2025}. 

The relationship between the characteristic bead ($\sim$150-500 km) and loop ($\sim$40-60 km) scales, separated by nearly an order of magnitude, has not been investigated until the present study. In Section \ref{sec:finestructure} we demonstrate that ViSP and VBI observations could be used in concert to bridge the gap between these two regimes.  Ribbon fragments with widths similar to the beads reported in other works are visible in both the VBI and ViSP data. The brightest beads in region r1A, initially hundreds of kilometers wide, appear to splinter into smaller fragments eventually 40--60 km wide (Figure \ref{fig:VBI_ultrafine}), the same spatial scale as the post-flare coronal loops observed by \cite{tamburri2025}. ViSP spectra (Figure \ref{fig:ViSP_ultrafine}) indicate that the decay phase ribbon splinters may be produced by downflowing material along post-flare coronal loops. 

We do not suggest that the observed ribbon fragmentation necessarily represents physical splintering of fundamental structures. During the flare impulsive phase, emission from actively flaring footpoints and loops may initially cause the ever-present background atmospheric structure to be indiscernible by contrast.  This phenomenon could be augmented by brief periods of adverse seeing, which would make smaller structures unresolvable. As the flare progresses, the bright bead-like topology dims, revealing smaller-scale, loop-like structures.  Figure \ref{fig:crossections} in Appendix \ref{sec:crosssections} shows that the bead-like structures are much more intense than the splinters that appear later, but that the brightness of the splinters still exceeds the pre-flare brightness of the same region. 

This interpretation is consistent with the findings of \cite{chitta2026}, who use flare data from the Sunrise III balloon mission \citep{korpi-lagg_2025} to diagnose the cause of fragmented ribbon structure. The authors suggest that uncombed, nonflaring chromospheric loops interspersed between actively flaring loops can produce the complex substructure observed in the flaring chromosphere. As also evidenced by their Figure 3 (similar to our Figure \ref{fig:slot_VBI}(c)), as the beads become less bright during the flare decay phase, the less bright non-flaring loops become more prominent, giving the impression of ``splintering."

Additionally, the number of individual peaks appears to be influenced by the Fried parameter (see Appendix \ref{sec:crosssections}). This suggests that bead-like structures hundreds of kilometers wide could often or ubiquitously be composed of smaller structures unresolvable due to impulsive flare emission, telescope resolution limits, or momentarily detrimental observing conditions. We note, however, that the initial stages of splintering around 22:34 UT occur during a \textit{decrease} in the Fried parameter, suggesting  that the bead splintering is not solely related to observing conditions.

There evidently remains an outstanding question in high-resolution flare physics: Do the coronal flare-time dynamics dictate flare fine structure, overwhelming the dynamics of the background active region chromosphere? Taken together, the ViSP and VBI data may support the notion that the bright fragments of a flare ribbon are composed of many coronal loops heated together and that some aspects of flare fine structure -- the apparent splintering of bright beads into smaller structures as the flare progresses, for example -- are introduced by flare dynamics.

Future analysis should determine whether the pre-flare atmosphere or flare-time processes (e.g., reconnection, tearing mode instability, modulation by SADs) are preferentially responsible for flare fine structure. For example, it would be informative to determine whether there is significant periodicity introduced by the flare that is not present in the pre-flare phase \citep[e.g.][]{french2021,french2025_beads}. Unfortunately, relatively poor atmospheric seeing conditions during the pre-flare phase obfuscate a detailed analysis of periodic structures throughout this flare. To further investigate the relevance of pre-flare structure to flare fine structure using ground-based observations, it will be necessary to capture a flare ribbon, including some period of pre-flare activity, with consistently good seeing.

Crucially, we note that the dynamics present in a single flare are not necessarily universal. \cite{ashfield2025_natas} present convincing evidence that modulation of electron acceleration in the corona impacts chromospheric flare fine structure.  Other studies (e.g., \citealt{yadav2025,french2021,french2025_downflows,french2025_beads}) demonstrate similar results. Given the variability in flare dynamics relevant to all layers of the solar atmosphere, and the massive amount of energy released by reconnection, it would be unsurprising if the non-flare chromosphere (including the presence of non-flare loops still present during the flare; \citealt{chitta2026}) overwhelms the substructure of the flare ribbon in one event, while coronal dynamics plainly imprint on the lower atmosphere in another. Relatedly, the beads observed on scales of 150--500 km in the past may still be produced by important physical processes, even if, as we suggest here, these beads are themselves composed of smaller structures currently only visible to the DKIST under the best observing conditions. We propose that a universal model for flare fine structure likely involves consideration of all possible mechanisms.

Our work demonstrates the importance of considering both high-resolution spectra and images in future flare models, as both reveal substantial fine structure on the smallest observable scales.  For example, in the likely scenario that the background plays a major role in the structure of the flare chromosphere, an accurate pre-flare atmosphere is a prerequisite for 1D flare modeling. Since the chromospheric structure varies on such small scales, the appropriate pre-flare atmosphere for individual flare loops may vary on equally small scales. In the past, however, models have been generated using a set of relatively few options for the initial atmosphere.  The development of accurate, pixel-specific pre-flare atmospheres is therefore an important challenge for future radiative-hydrodynamic flare models. 

\section{Conclusions}\label{sec:concc6}

In this study, we have presented high-resolution, high-cadence images and spectroscopic observations of a C-class flare from DKIST.  We investigate the spectral complexity of the chromosphere during the flare impulsive phase and demonstrate the multiscale nature of ribbon fragments in both images and spectra.  Our main findings are as follows:

\begin{enumerate}
    \item ViSP and VBI observations of the lower solar atmosphere, primarily the chromosphere, indicate a complex spectral structure and highly fragmented flare ribbons.
    \item \textit{k}-means clustering of the ViSP Ca \textsc{II} H and H$\beta$ spectra reveals four spectral types during the flare: (i) redshifts, (ii) symmetric, unshifted spectra, (iii) central reversals, and (iv) blueshifts.  The spatial variability is most clear in the Ca \textsc{II} H line. By clustering the spectra from many ViSP scans, we parameterize the spectral dynamics of the flare ribbon in a physically meaningful way.  The results of the clustering technique could serve as valuable constraints for future dynamic flare models.
    \item VBI images and ViSP intensity maps reveal that flare ribbons appear to be composed of localized regions of heating $\sim$300--470 km wide.  Notably, during the impulsive and early decay flare phases between 22:33 UT and 22:37 UT, a single flare bead \textit{appears} to splinter into smaller structures on timescales of a few minutes as flare emission wanes.  Eventually, small loop-like structures are visible in the locations where the beads are originally observed.  Our work suggests that ribbon beads may be composed of smaller, loop-like structures in the lower corona 40--60 km wide, visible only during periods of better seeing.  This is the same scale as reported in post-flare loop observations by \cite{tamburri2025}.  
    \item Spectroscopic observations of the ribbon beads confirm that small-scale ribbon fragments correspond to emission in the red wing of chromospheric lines (specifically Ca \textsc{II} H), supporting the notion that the splinters in the decay phase of the ribbon evolution are plasma downflows along post-flare coronal loops. The evolution of these structures -- from initial brightening to decay-phase splintering -- could provide clues for future studies of the thermodynamic evolution of individual heating events.
\end{enumerate}

It is not yet known whether modulation by reconnection, the structure of the background chromosphere, or some combination of both are responsible for near-ubiquitous spectral and spatial variation in flare ribbon fine structure. Simultaneously, a comprehensive flare model explaining all spectral dynamics and evaluating all potential heating mechanisms continues to elude the field. 

To address these remaining mysteries, this work demonstrates that DKIST has a unique capability to track the essential fine-scale spatial and spectral properties of ribbon beads, loops, and other structures at very small scales. Might the highly structured flare chromosphere be produced by spatial variation in heating scenarios? It is feasible, for example, that the brightest features could be generated by one mechanism (a high-energy beam of accelerated particles, for example, evidenced by the temporal correlation between hard X-ray signatures and the brightening of region r1A) while the remainder of the ribbon is heated by some other population of particles or a different mechanism altogether.  Since such non-uniformity in heating could be important down to spatial scales at or below the limits of modern observational capabilities, it is crucial to \textit{simultaneously} study the spectral diversity and spatial fine structure of solar flares. Both provide essential diagnostics that will inform future state-of-the-art flare models.  To develop these models and deepen our understanding of flare fine structure, future high-resolution observations of flares with DKIST are imperative, ideally together with observations from Solar Orbiter, IRIS, Solar-C (e.g., \citealt{suematsu2016}), the Multi-Slit Solar Explorer (MUSE; \citealt{cheung2022,depontieu2022}), SST, and others.

\begin{acknowledgments}

The research reported herein is based in part on data collected with the Daniel K. Inouye Solar Telescope (DKIST), a facility of the U.S. National Science Foundation (NSF). DKIST is operated by the National Solar Observatory under a cooperative agreement with the Association of Universities for Research in Astronomy, Inc. (AURA). DKIST is located on land of spiritual and cultural significance to Native Hawaiian people. The use of this important site to further scientific knowledge is done with appreciation and respect. Support for this work was provided by the NSF through the DKIST Ambassadors Program, administered by the National Solar Observatory and AURA. 
This work was also supported by the NSF Astronomical Sciences program through grant 2407849. 
CAT and AFK acknowledge funding support from NSF award 2407850. CAT acknowledges support from the
Science and Technology Facilities Council (STFC) under grant No. ST/X000923/1.
RJF acknowledges support from NASA HGI award 80NSSC25K7927. GSK acknowledges support from a Royal Society University Research Fellowship (URF\textbackslash R1\textbackslash 251160). We thank the anonymous referee for their helpful commentary, which has greatly improved the presentation of this work. 
    
    \end{acknowledgments}
    
\bibliography{refs}
\bibliographystyle{aasjournal}

\appendix

\section{Determining the optimal number of clusters $k$}\label{app0}

\begin{figure*}
    \begin{center}\includegraphics[width=1\linewidth]{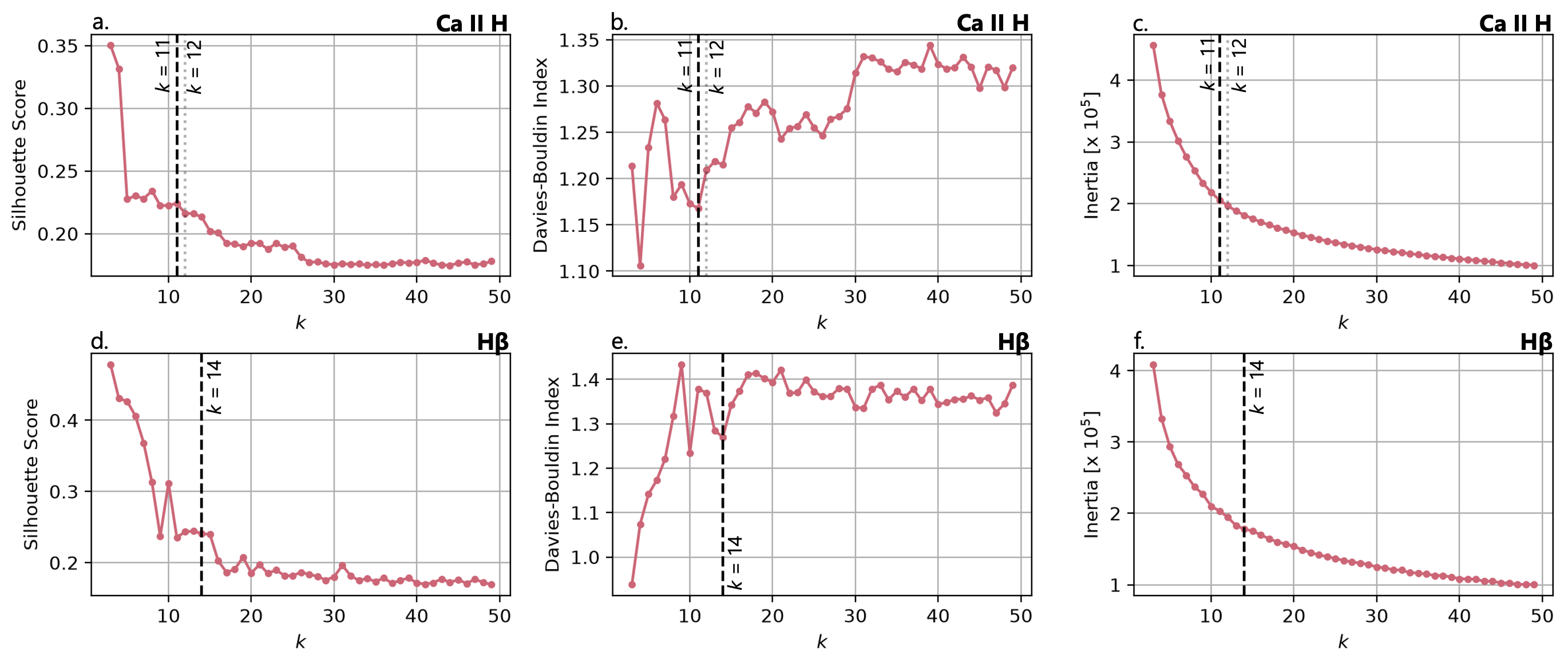}
    \end{center}
    \caption{Performance metrics used to determine the optimal number of clusters in Section \ref{sec:kmeans}.  (a)-(c) Silhouette score, Davies-Bouldin index, and inertia for the Ca \textsc{II} H clustering, $k=$ 3 to 49.  (d)-(f) The same, for H$\beta$ clustering. Vertical dashed lines indicate the optimal number of clusters based on the three metrics.  The vertical dotted line in (a)-(c) indicates the choice of $k$ used in the text after inspecting the fine structure of clustered spectra.}
    \label{fig:clustering metrics}
\end{figure*}

To justify our choice of the number of clusters $k$ in Section \ref{sec:kmeans}, we interrogate three metrics: (1) the silhouette score (which measures intra-cluster cohesion and inter-cluster separation; higher is better); (2) the Davies-Bouldin index (which measures intra-cluster dispersion and inter-cluster separation; lower is better); and (3) inertia (which measures how compact a cluster is – how close the members are to the cluster center – most commonly used in the \textit{elbow method}). Determining the optimal number of clusters is often difficult especially in higher-dimensional cases such as this. We identify the values for $k$ for which all three metrics best agree, shown in Figure \ref{fig:clustering metrics}.  While $k=10$ has a local peak in silhouette score and minimum in Davies-Bouldin index for the H$\beta$ clustering, the inertia continues to decrease substantially for higher $k$.  We therefore choose $k=14$ for H$\beta$, where the inertia curve begins to flatten and the other two metrics are also favorable. The optimal number of clusters based on these three scores is $k=11$ for Ca \textsc{II} H and $k=14$ for H$\beta$.

None of the metrics present indisputable evidence in favor of a specific value for $k$.  The inertia curves, for example, lack a clear elbow, as is often the case for high-dimensional clustering. We therefore inspect the results of the clustering algorithm for values of $k$ near (but greater than) the optimal values determined above. We find that increasing the number of Ca \textsc{II} H clusters to $k=12$ adds a grouping that correctly identifies the smallest features visible to ViSP (dashed and solid boxes, Figure \ref{fig:kmeans_manyscan_caii}(a)), and therefore report the $k=12$ results in Section \ref{sec:kmeans}. Similar adjustments do not yield appreciably different results for H$\beta$. Small differences in the choice of $k$ do not affect our identification of the four main spectral categories discussed in the text.

\section{Comparison of $k$-means results to observed spectra}\label{app1}

\begin{figure*}
    \begin{center}\includegraphics[width=.85\linewidth]{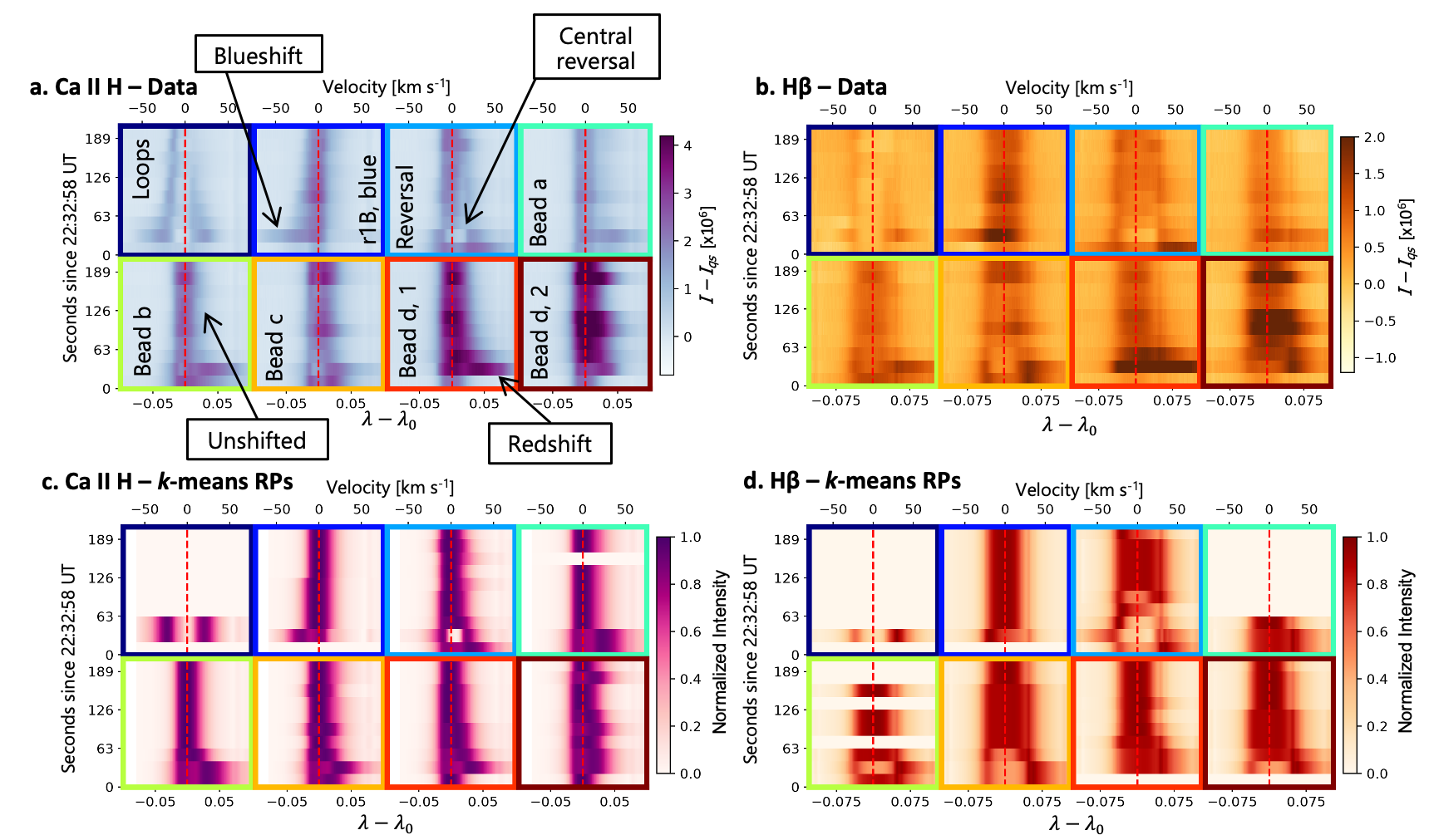}
    \end{center}
    \caption{Comparison of observed ViSP spectra and corresponding RPs from $k$-means clustering for select pixels. In (a) and (b) we show the Ca \textsc{II} H and H$\beta$ spectra, respectively, for the ViSP scans from 22:32:58 UT to 22:36:07 UT at the colored crosses in Figure \ref{fig:ViSP_evolve}(a--c) and (g--i).  Border colors for each panel correspond to the cross color in Figure \ref{fig:ViSP_evolve}.  In panel (a) we note the main spectral profile types identified here.  In panels (c) and (d) we show the Ca \textsc{II} H and H$\beta$ RPs from the \textit{k}-means analysis, respectively, for the same pixels as are analyzed in panels (a) and (b). In panels (a) and (b), the intensity in the second scan of the ViSP is reduced by a factor of 1.5 for clarity.}
    \label{fig:ViSP_datavskmeans_evolve}
\end{figure*}

Although the $k$-means clustering in Section \ref{sec:kmeans} is performed using intensity-normalized spectra, the algorithm captures the time evolution of spectra remarkably well.  In Figure \ref{fig:ViSP_datavskmeans_evolve}(a--b) we show the evolution of Ca \textsc{II} H and H$\beta$ spectra in select pixels (locations indicated in Figure \ref{fig:ViSP_evolve}(a--c) and (g--i)) demonstrating the four primary spectral categories. Figure \ref{fig:ViSP_datavskmeans_evolve}(c--d) shows the time evolution of RPs for the same pixels from the \textit{k}-means analysis for each of the ten scans analyzed (from 22:32:58 UT to 22:36:07 UT). All four spectral types are visible in Ca \textsc{II} H and H$\beta$, but especially in the former. Occasionally a pixel is too faint to be reliably included in the flare mask without also including much of the non-flare chromosphere in the mask.  No RP is shown in these cases.

This demonstrates the effectiveness of the $k$-means clustering in characterizing the spatiotemporal evolution of the ViSP spectra despite the far fewer number of RPs compared to ViSP pixels, a useful tool for guiding future flare modeling.

\section{Time evolution of ribbon fragmentation in relation to Fried parameter}\label{sec:crosssections}

\begin{figure*}
     \begin{center}
    \includegraphics[width=.8\linewidth]{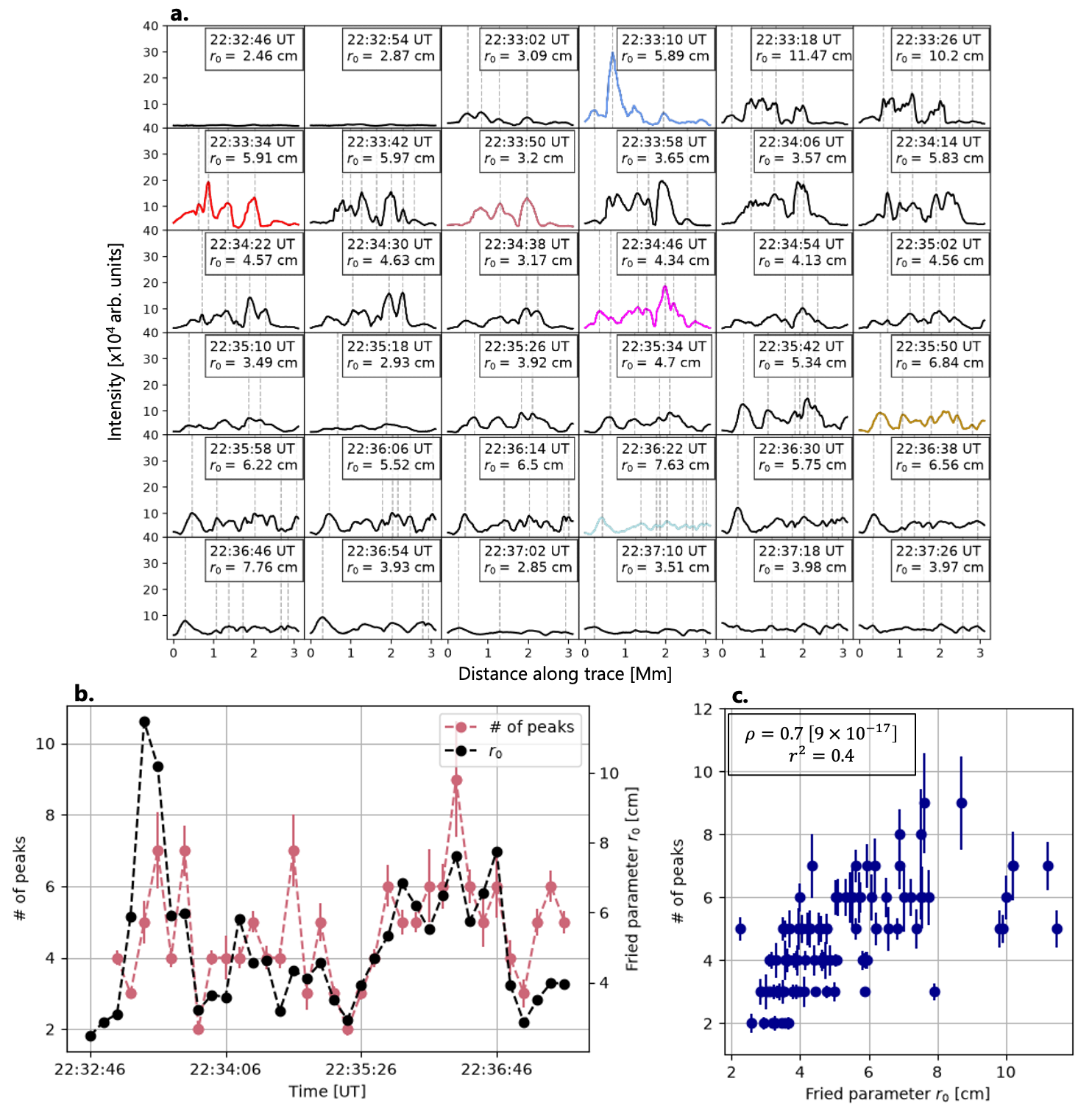}
    \end{center}
    \caption{Evolution of fine structure in ribbon r1A over five minutes.  (a) Intensity along the ribbon traces in Figure \ref{fig:slot_VBI}(c) at every third time step between 22:32:46 UT and 22:37:26 UT. Vertical lines indicate the positions of detected peaks. The color of select intensity profiles corresponds to the vertical lines in Figure \ref{fig:slot_VBI}(c). (b) The number of peaks detected in the ribbon trace, compared to the Fried parameter, in each panel of (a). (c) A comparison of the Fried parameter $r_0$ to the detected number of peaks in the ribbon cross-section of every frame between 22:32:46 UT and 22:37:26 UT. We report the Spearman correlation coefficient $\rho$ (with associated p-value) and Pearson correlation coefficient $r^2$ in the inset box in (c).}
    \label{fig:crossections}
\end{figure*}

We further analyze the flare ribbon trace defined in Section \ref{sec:trace} to more rigorously identify peaks in the ribbon intensity.  Using the \texttt{find\_peaks} function in the \texttt{scipy.signal} Python package, we identify the number of peaks in the ribbon trace intensity at the times shown in Figure \ref{fig:crossections}(a).  We normalize the intensity array along the trace in each time step and set the \texttt{prominence} parameter (the intensity difference between a peak and the higher of the two minima on either side of the peak) in \texttt{find\_peaks} to 0.05. We exclude peaks for which the error in FWHM exceeds 50\%.  In Figure \ref{fig:crossections}(b) we plot the identified number of peaks and the Fried parameter at the time steps in panel (a). We approximate the uncertainty in the number of peaks based on the median uncertainty in fragment width, assuming that the number of peaks is inversely proportional to the median peak width in the slice: $\delta N= N \cdot\operatorname{med}({\frac{\delta w_i}{w_i}})$ where $w_i$ is the width of each detected peak $i$ and $\delta w_i$ is the standard deviation error in $w_i$ calculated from \texttt{scipy.optimize.curve\_fit}.

Under better observing conditions (e.g., around 22:33:26 UT or between 22:35:26 UT and 22:37 UT) there are more detected peaks.  In panel (c) we compare the number of peaks to the Fried parameter in each frame between 22:32:46 UT and 22:37:26 UT.  Although the squared Pearson's correlation coefficient ($r^2=0.4$) measuring the linear relationship between the two quantities is moderate, the Spearman correlation coefficient ($\rho=0.7$) is high and significant (significance value $p\ll0.01$), suggesting that the Fried parameter and number of identified peaks are positively associated.  There are likely to be many other factors influencing the observed number of ribbon fragments -- see Section \ref{sec:disc_c6}. For the few frames with moderate to above average observing conditions ($r_0 \gtrsim 7$ cm), the relation is far weaker. This suggests that above a certain threshold, either the observed strands no longer decrease in size, or the diffraction limit of the telescope is insufficient to observe existing smaller structures. 

\end{document}